\documentclass[12pt]{article}

\makeatletter
\renewcommand\@biblabel[1]{#1.}
\makeatother

 \usepackage[superscript]{cite}
 \usepackage{setspace}
\usepackage[english]{babel}
\usepackage[utf8]{inputenc}
\usepackage{amsmath, amsthm, amssymb,amsfonts,bm}
\usepackage{tikz}
\usetikzlibrary{positioning,shapes.geometric,graphs}
\usetikzlibrary[graphs]
\usepackage{graphicx}
\usepackage{placeins}
\usepackage{color}
\usepackage{multirow}
\usepackage{soul}
\usepackage{url}
\usepackage{verbatim}
 \usepackage[english]{babel}
\usepackage[utf8]{inputenc}
\usepackage{amsmath, amsthm, amssymb,amsfonts}
\usepackage{tikz}
\usetikzlibrary{positioning,shapes.geometric,graphs}
\usetikzlibrary[graphs]
\usepackage{color}
\usepackage{verbatim}
\usepackage{lscape}
\usepackage{hyperref}
\usepackage{float}
\usepackage{comment}
\usepackage{mathrsfs}
\usepackage{multicol}
\usepackage{bbm}
\usepackage{lipsum}
  \usepackage{color}
\usepackage{subcaption}  
\usepackage{float}
\usepackage{comment}
\usepackage{mathrsfs}
\usepackage{multicol}
\usepackage{lipsum}
 \usepackage{setspace}
 \usepackage{longtable}
\usepackage{geometry}
\newenvironment{myfont}{\fontfamily{phv}\selectfont}{\par}  
\DeclareTextFontCommand{\textmyfont}{\myfont}

\title{Estimating Individualized Treatment Rules in Longitudinal Studies with Covariate-Driven Observation Times \vspace{0.5cm}\\
\small{Janie Coulombe\footnote{Department of Epidemiology, Biostatistics and Occupational Health, McGill University}, Erica E. M. Moodie\footnote{Department of Epidemiology, Biostatistics and Occupational Health, McGill University}, Susan M. Shortreed\footnote{Biostatistics Unit, Kaiser Permanente Washington Health Research Institute, Seattle, Washington; Biostatistics Department, University of Washington}, Christel Renoux\footnote{Lady Davis Institute for Medical Research, Jewish General Hospital, Montreal; Department of Neurology and Neurosurgery, McGill University; Department of Epidemiology, Biostatistics and Occupational Health, Mcgill University}}\\
\date{}}

\renewcommand{\arraystretch}{3}

\begin{document}
\maketitle

\begin{abstract}
The sequential treatment decisions made by physicians to treat chronic diseases are formalized in the statistical literature as dynamic treatment regimes. To date, methods for dynamic treatment regimes have been developed under the assumption that observation times, i.e., treatment and outcome monitoring times, are determined by study investigators. That assumption is often not satisfied in electronic health records data in which the outcome, the observation times, and the treatment mechanism are associated with patients' characteristics. The treatment and observation processes can lead to spurious associations between the treatment of interest and the outcome to be optimized under the dynamic treatment regime if not adequately considered in the analysis. We address these associations by incorporating two inverse weights that are functions of a patient's covariates into dynamic weighted ordinary least squares to develop optimal single stage dynamic treatment regimes, known as individualized treatment rules. We show empirically that our methodology yields consistent, multiply robust estimators. In a cohort of new users of antidepressant drugs from the United Kingdom's Clinical Practice Research Datalink, the proposed method is used to develop an optimal treatment rule that chooses between two antidepressants to optimize a utility function related to the change in body mass index.
\end{abstract}

\section{Introduction}

In recent years, significant effort has been put towards developing statistical methods that can leverage observational data to make valid causal inference about treatment (or exposure) effects (see e.g., Bang and Robins\cite{bang2005doubly}, Stuart\cite{stuart2010matching}, Moodie et al. \cite{moodie2012q}, and Schuler and Rose\cite{schuler2017targeted}). Much of the literature on causal inference has focused on assessing marginal effects of treatments in the whole population, i.e., how the outcomes in the study population would differ, on average, had we given everyone one treatment versus another. Though such marginal effects are often interesting from a policy standpoint, they are not always the most relevant in clinical practice, where patients hope to receive a treatment that is tailored to their unique characteristics. Individualized treatments (often falling under the umbrella term of \textit{precision medicine} \cite{kosorok2019precision}) may be especially interesting in settings where it is known that the treatment effect for some individuals differs considerably from the marginal effect. In this work, we focus on the estimation of dynamic treatment regimes (DTRs), which formalize individualized, possibly sequential, treatment decisions taken as functions of patient's characteristics. We limit the DTRs under consideration to those that optimize an expected outcome, so called `optimal' DTRs. 

Three common methods for developing optimal DTRs are g-estimation \cite{robins1992g}, Q-learning (see Laber et al.\cite{laber2014q} for a review), and dynamic weighted ordinary least squares (dWOLS) \cite{wallace2015doubly}. These methods are implemented in standard software (see e.g., Wallace et al.\cite{wallace2017dynamic,wallace2016package}, Simoneau et al.\cite{simoneau2020optimal}, Tsiatis et al.\cite{tsiatis2019dynamic}, Linn et al.\cite{linn}, McGrath et al.\cite{mcgrath2020gformula}). The latter method, dWOLS, optimizes the expected outcome by estimating treatment effects within strata of patients' characteristics; these effects are sometimes termed \textit{effect modifications} by patients' characteristics. The dWOLS method provides intuitive estimators for optimal DTRs while combining the advantages of Q-learning and propensity-score based weighting \cite{horvitz1952generalization,rosenbaum1987model}. It also achieves similar properties to the estimators derived from g-estimation but under a more familiar framework. In particular, under conditions stated in Section 2, the estimators derived using dWOLS are doubly-robust in the sense that they are consistent if either the treatment model or the outcome model is correctly specified.  

Electronic health records (EHR) data are often used to develop DTRs.\cite{johnson2018causal, simoneau2020optimal,coulombe2021can} These data are recorded irregularly across all patients, with the observation of patients' outcome and treatment likely depending on their unique characteristics (such as symptoms, comorbidites, age, sex, etc.). That is, observation indicators, which represent whether or not a patient was observed at a given time, are associated with covariates that could be associated with the longitudinal outcome. In such situations, conditioning on observed data without making any adjustments for the treatment and observation processes may lead to DTRs with spurious associations caused by collider-stratification bias \cite{greenland2003quantifying}. The observational nature of studies that are based on EHR data also means that patients were not randomized to treatment but prescribed treatment based on their individual characteristics (a feature that is commonly called \textit{confounding} when the same covariates affect the longitudinal outcome). Given the spurious associations mentioned above, using EHR data to assess treatment effects (or treatment effect modifications) must be done carefully. Drawing a causal diagram that represents the assumed data generating mechanism can help in determining whether these associations are problematic \cite{coulombe2021estimation}.

Though the statistical literature has extensively discussed confounding in the context of developing optimal DTRs, it has paid little attention to covariate-driven observation times. Robins et al. \cite{robins2008estimation} discussed the identification of causal effects when jointly modelling DTRs and monitoring (observation) schedules as well as some issues related to the extrapolation of optimal treatment and testing strategies. They introduced a \textit{no direct effect} assumption for observation, which requires that observation decisions have no effect on patient characteristics (including outcomes) after conditioning on the treatment decision. Neugebauer et al. \cite{neugebauer2017identification} extended their work to settings with survival outcomes and differentiated five classes of counterfactual variables that may be of interest in such context. Kreif et al.  \cite{kreif2018evaluation} tackled another important issue related to the varying observation schedules in EHR data and the development of DTRs by proposing a way to analyze irregularly measured time-varying confounders \cite{kreif2018evaluation}. Bayesian approaches have been proposed to estimate optimal two-stage strategies in settings with interval censoring \cite{thall2007bayesian} and to estimate causal effects via g-computation under irregular observation schedules \cite{shahn2019g}. Most recently, Yang\cite{yang2021semiparametric} proposed a methodology to estimate the parameters of a continuous structural nested mean model when data are irregularly observed and the observation schedule may confound the treatment effects. Yang built on the work of Lok\cite{lok2017mimicking}, who had proposed estimating equations for the same parameters. An important development of Yang is the use of semiparametric theory of influence functions to construct an efficient estimator for continuous-time structural nested mean models. The method does not model explicitly the observation times and relies on a no unmeasured confounding assumption based on a martingale condition, which does not allow for mediators of the treatment-outcome relationship to drive observation times.  

While few methods have been proposed in the literature on DTRs that simultaneously account for covariate-driven mechanisms for treatment and observation times, the issues mentioned above have been covered in the literature on the estimation of the marginal effect of covariates (e.g., treatment) on a longitudinal outcome (see, e.g., Goldstein et al. \cite{goldstein} who demonstrated the strength of bias due to the association of a risk factor with the outcome and the observation processes, and McCulloch et al.\cite{mcculloch} who also discussed the estimation of such marginal parameters when the corresponding covariates are associated with random effects). To account for the observation process, authors have proposed the use of inverse intensity of visit (\textmyfont{IIV}) weights \cite{lin2004analysis, buuvzkova2009semiparametric,zhu2017estimation}, random or latent effects \cite{liang2009joint,cai2012time,dai2018joint}, or fully parametric inference by specifying the full joint likelihood of the outcome and observation processes \cite{lipsitz2002parameter}. Within the causal inference framework, the bias due to the spurious associations mentioned above in the estimation of the marginal effect of a binary treatment on a longitudinal outcome using EHR data was demonstrated in Coulombe et al. \cite{coulombe}. In that work, two semiparametric estimators were proposed for the causal marginal effect of a binary treatment on a continuous longitudinal outcome that accounted for the covariate-driven treatment and observation mechanisms \cite{coulombe}. Here, we extend one of these estimators to the case of DTRs. 

\textcolor{black}{In this work, we focus on a single stage rule, known as an individualized treatment rule (ITR), which is a special case of a DTR. We consider repeated measurements of the treatment and outcome of each individual. Patients can, therefore, contribute multiple measurements in the estimation of the ITR, which we term a \textit{repeated measures ITR}.} The more general case of DTRs comprising multiple sequential rules corresponding to multiple treatment decisions is a topic of future work. We show here that by extending one of the estimators proposed in Coulombe et al. \cite{coulombe} to an ITR setting, we can consistently estimate the conditional effect of treatment within strata of patients' variables rather than a single marginal effect. For that, we use dWOLS and a new weighting mechanism that incorporates independently the informative observation times and the treatment process under assumptions that are commonly postulated in the causal inference literature. To our knowledge, we propose the first estimator for an optimal repeated measures ITR for binary treatment and continuous longitudinal outcome that applies to data subject to covariate-driven observation and treatment processes.

 This paper is divided as follows. We introduce the proposed methodology and the required assumptions in Section 2. We test the method through extensive simulation studies, the details and results of which we describe in Section 3. In Section 4, we apply the methodology to develop a repeated measures ITR that chooses between two commonly prescribed antidepressants, citalopram and fluoxetine, to maximize a utility function related to changes in body mass index (BMI). The optimal treatment rule is estimated using a cohort of patients with depression taken from the United Kingdom's (UK) Clinical Practice Research Datalink (CPRD)\cite{herrett2015data}. Finally, a discussion follows in Section 5.

\section{Methods}

\subsection{Notation}

We suppose that we have a random sample of $n$ individuals, taken from a larger population, that is indexed by $i=1,...,n$. We use bold notation to refer to both vectors and matrices. 
For each individual in the population, \textcolor{black}{we are interested in the estimation of a repeated measures ITR that optimizes a continuous outcome denoted by $Y_i(t)$. By ITR, we mean a one-stage treatment rule that does not require optimization over several time points simultaneously but rather searches for the ``cross-sectional'' treatment rule that, at a given point, optimizes the outcome. By repeated measures, we mean that patients can contribute multiple observations in the estimation of the ITR (where each observation is a vector containing a treatment value, an outcome value, and so on). We discuss in the next paragraphs the assumptions required about the treatment effect to ensure that such repeated measures ITR can be estimated consistently.} The treatment rule is based on the estimated effect modification by some covariates (patients features) that we call \textit{tailoring variables} and that we denote by $\mathbf{Q}_i(t)$. The treatment $A_i(t)$ is binary and takes values in $\left\{0, 1\right\}$, and both the treatment and the continuous outcome $Y_i(t)$ can vary over time $t$.   

Assume that a larger $Y_i(t)$ is better, such that we aim for a treatment rule that maximizes $Y_i(t)$. The outcome is assumed to be measured irregularly across patients, at individual-specific times $T_{i1},...,T_{iF_i}$ contained in $[0,\tau]$ with $\tau$ the maximum follow-up time across the study cohort. Patients are, therefore, allowed to have a different number of visits $F_i$ and varying gap times between their visits. The observation (or monitoring) indicators $dN_i(t)$ are equal to 1 if individual $i$ was observed at time $t$, and 0 otherwise. These indicators can be seen as part of a counting process $N_i(t)$ that counts the number of observation times (visits) by time $t$, with $N_i(t) = \int_{s=0}^t dN_i(s) = \int_{s=0}^ t \sum_{j=1}^{F_i} \mathbb{I}(s= T_{ij})$. While the outcome is assumed to be observed at those individual-specific times, the treatment and all covariates that will be used in nuisance models (observation and treatment models) are assumed to be measured continuously in time and the tailoring variables $\mathbf{Q}_i(t)$ are assumed to be measured at least at the same times as the treatment, and possibly at other times too. That assumption is not unrealistic, especially for covariates related to prescription drugs, which are generally recorded in EHR automatically at the time of prescription. \textcolor{black}{An example of such an observation setting is a study using EHR data in which the study treatment is defined using drug prescriptions, always available to the data analyst, in which the outcome is measured sporadically (e.g., a weight or a blood pressure outcome) and in which we are interested in effect modification by sex, our tailoring variable}. We denote by $\mathbf{K}_i(t)$ the set of potential confounders for the causal effect of $A_i(t)$ on $Y_i(t)$ and by $\mathbf{V}_i(t)$ a complementary set of covariates that can affect or be associated with the observation indicator $dN_i(t)$. We broadly assume that covariates in $\mathbf{V}_i(t)$ are related to the treatment $A_i(t)$, the outcome $Y_i(t)$, or both (possibly inducing biasing associations for the causal effect of interest). We denote by $\mathbf{Z}_i(t)$ the variables in $\mathbf{V}_i(t)$ that mediate the effect of treatment $A_i(t)$ on $Y_i(t)$. Tailoring variables $\mathbf{Q}_i(t)$ are allowed to contain confounders or pure predictors of the outcome, and they may share variables with $\mathbf{V}_i(t)$ if these variables affect the observation times (although $\mathbf{Q}_i(t)$ should not contain mediators of the treatment effect on the outcome, to avoid bias in the estimation of effect modifications). The assumed data generating mechanism is depicted in Figure \ref{fig2}.

 \subsection{Assumptions}
 
 \textcolor{black}{Some assumptions on the acuteness of the treatment effect are required for using a repeated measures ITR. Omitting the patient index for ease of notation, we broadly assume that $A(t)$ is the treatment associated with the outcome $Y(t)$ over what we call a treatment interval for the outcome $Y(t)$. The treatment $A(t)$ may have been measured (observed to be prescribed) previous to time $t$, and it is assumed that the corresponding outcome $Y(t)$, if $dN(t)=1$,  is observed at a time when treatment $A(t)$ had the time to be effective. It is, therefore, necessary to have a treatment with an effect that is acute enough (and not too much delayed) such that any observed outcome corresponding to a particular treatment interval is only affected by the treatment corresponding to that treatment interval. That is, there should be no treatment effects overlap across treatment intervals, nor delayed effects of treatment, nor treatment interactions across different treatment intervals. In studies where the gap times between observation times for the outcome are very large, treatment intervals may span longer periods of time, and the acuteness assumption may need to be relaxed. While relaxing the acuteness assumption is of interest, it is out of the scope of the current paper and, going forward, we assume that the acuteness assumption holds.}
 
 We use the potential outcome framework\cite{neyman1923application,rubin1974estimating} and introduce two potential outcomes for inference, denoted by $Y_{i}^1(t)$ and $Y_{i}^0(t)$, where $Y_{i}^1(t)$ represents the outcome we would observe for individual $i$ at time $t$, had they received treatment $1$ over the corresponding treatment interval as defined above, and $Y_{i}^0(t)$, had they received treatment $0$. Only one of these two potential outcomes can actually be observed at a given time, as the binary treatment assumption implies an individual can only receive one of the two treatment options. 
 
 For simplicity, denote by $\mathbf{X^{\beta}}(t)$ the matrix of risk factors for the outcome, which in our case is composed of columns $\mathbf{K}(t)$, $\mathbf{Q}(t)$, and a first column of $1$s for modelling the intercept ($\mathbf{X^{\beta}}(t)$ should \textit{not} contain any mediator of the treatment's effect on the outcome). Sets $\mathbf{K}(t)$ and $\mathbf{Q}(t)$ could be entirely different or exactly the same, as would be the case if the tailoring variables are the confounders, and shared columns between the two sets should not be included twice in building matrix $\mathbf{X^{\beta}}(t)$. Further denote by $\mathbf{X^{\psi}}(t)$ the matrix comprising the tailoring variables (that we also denoted by $\mathbf{Q}(t)$). Suppose that the full matrix $\mathbf{X}(t)$ is such that
$$\mathbf{X}(t)=\left[  \mathbf{X^{\beta}}(t) \hspace{0.2cm} \mathbf{X^{\psi}}(t)  \right].$$  
 As in Coulombe et al. \cite{coulombe}, we make the following assumptions (P1)-(P3):
  \begin{align}
  \tag{P1} A_i(t) \perp \left\{Y_{i}^0(t), Y_{i}^1(t)\right\}  |  \mathbf{K}_i(t), \mathbf{V}_i(t), dN_i(t) \hspace{0.2cm}\text{(conditional exchangeability),}\label{p11}\\
  \tag{P2} 0  < \mathbb{P}(A_i(t)=0|  \mathbf{K}_i(t) ), \mathbb{P} (  A_i(t) =1|  \mathbf{K}_i(t) )  <1  \hspace{0.2cm}\text{(positivity of treatment), and}\label{p12}\\
  \tag{P3} Y_{i}^a(t)=Y_i(t) \hspace{0.1cm}\text{if}\hspace{0.2cm} A_i(t) =a\hspace{0.2cm}\text{(consistency of the outcome)}.\label{p13}
    \end{align}

\noindent   Assumptions  (\ref{p12}) and (\ref{p13}) are standard in all causal inference settings while (\ref{p11}) is specific to the setting with covariate-driven observation times. Assumption (\ref{p11}) means that upon conditioning on the observation indicator $dN_i(t)$ (i.e., keeping only the observed outcomes in the analysis), the sets of covariates $\mathbf{K}_i(t)$ and $\mathbf{V}_i(t)$ are together sufficient to break any potential biasing associations for the causal effect of interest.
  
  To account for the observation process, we assume that observation at time $t$ depends on set $\mathbf{V}_i(t)$ and that the observation intensity can be modelled by a proportional rate model as follows
\begin{align}
\tag{V1} \mathbb{E}[dN_i(t)| \mathbf{V}_i(t)] =\xi_i(t)\exp\left\{ \bm{\gamma}' \mathbf{V}_i(t) \right\} d\Lambda_0(t),\label{ratee2}
\end{align}
where $\xi_i(t)=\mathbb{I}(C_i \ge t)$ is an indicator of being at risk at time $t$, with $C_i$ the censoring time of individual $i$, i.e., the time when an individual is lost to follow-up. The function $\Lambda_0(t)$ is any non-decreasing function \cite{lawless1995some, lin2001semiparametric}. The proportional rate model in (\ref{ratee2}) allows for the observation times to occur at any time (continuously) and irregularly across patients, as a function of their characteristics. In this work, we assume that censoring is uninformative after conditioning on the treatment, the tailoring and the confounder variables, which can be expressed as
\begin{align}
\tag{C1} \mathbb{E}[Y_i^a(t) |  \mathbf{X}_i(t), C_i \ge t] = \mathbb{E}[Y_i^a(t) |  \mathbf{X}_i(t)].
\end{align}
This assumption can be extended to the setting where censoring is informative by additionally using inverse probability of censoring weights (see e.g., Robins et al.,\cite{robins2008estimation} Section 3). Finally, we also assume positivity of observation, an assumption that is denoted by 
\begin{align}
\tag{V2} 0 < \mathbb{E}[dN_i(t)| \mathbf{V}_i(t)] <1 \hspace{0.2cm} \forall t.\label{lablab}
\end{align}
 Times at which the probability of observing an individual is zero should not be included in the analysis set (for example, times when a patient is not yet enrolled in the health system). These times would not only preclude treatment positivity (as the treatment would not be allowed to change), but they could also lead to bias in the causal estimation due to extrapolations in regions of the domain of time when a patient had no chance of being observed.

The interactions between the treatment and tailoring variables and their effects on the mean outcome can inform the best treatment decision to maximize an expected outcome $\hat{Y_i}(t)$. We, therefore, base the ITR on those interactions. Optimizing (in this case, maximizing) the expected outcome is our ultimate goal in tailoring the treatment to the individual. While the more commonly estimated causal marginal effect is a population-average effect, the ITR we aim to estimate is based on a conditional outcome mean model that is used to estimate the treatment effects within strata of the tailoring variables.  
 
The following outcome mean model is further postulated, conditional on the risk factors and tailoring variables:
 \begin{align}
\tag{O2} \mathbb{E}[Y_i(t)| A_i(t), \mathbf{X}_i(t)] = f \left\{\mathbf{X}_i^{\beta}(t) ; \bm{\beta} \right\}   +  A_i(t)\bm{\psi}'   \mathbf{X}^{\psi}_i(t).\label{outcomee}
\end{align}
 
The first term in (\ref{outcomee}), $f\left\{ \mathbf{X}_i^{\beta}(t) ; \bm{\beta} \right\}$, is called the \textit{treatment-free} model and is a function of the risk factors for the outcome. The second term comprises the treatment indicator, $A_i(t)$, and the blip function, $\bm{\psi}'   \mathbf{X}^{\psi}_i(t)$. \textcolor{black}{As briefly mentioned earlier, we must assume that treatment effects are acute enough not to overlap across treatment intervals, and that there are no synergistic or antagonistic effects between any subsequent treatments of a patient (i.e., treatments $A_i(s)$ and $A_i(t)$ for $s<t$)\cite{larry}. These conditions ensure that the ITR can be estimated consistently using repeated measurements of the same individual, without any carryover treatment effect that could bias the ITR.} Under all conditions stated above and if the model in (\ref{outcomee}) represents the true outcome generating mechanism, the blip function indicates how the outcome varies when going from treatment 0 to treatment 1 (that is, the difference between the two potential outcomes). In particular, the outcome mean is larger under $A_i(t)=0$ if $\bm{\psi}'  \mathbf{X}^{\psi}_i(t)  < 0$, and conversely, larger under $A_i(t)=1$ if $\bm{\psi}'  \mathbf{X}^{\psi}_i(t) \ge 0$. Therefore, by estimating the blip function, one can determine which treatment should be prescribed to optimize the expected outcome. Note that in situations like our motivating example, where there are two active treatments, this model is not an expected outcome ``in the absence of treatment'' but rather at the reference level of treatment.

\subsection{Proposed methodology}

To estimate an optimal ITR under our postulated assumptions, it suffices to estimate the coefficients $\bm{\psi}$ in (\ref{outcomee}). This can be done using dWOLS \cite{wallace2015doubly} which, in our setting, corresponds to a weighted least squares regression because no optimization over time points is required \cite{larry}. The dWOLS method leads to estimators  $\hat{\bm{\psi}}$ for optimal treatment rules of the form
\begin{align}
\text{ ``Treat with $A_i(t)=1$ if $\hat{\bm{\psi}}'  \mathbf{X}^{\psi}_i(t) \ge 0$, and with $A_i(t)=0$ otherwise''.}  \label{otr}
\end{align}
If one or both of the treatment model and the outcome model are correctly specified, the blip function is correctly specified, and we use an inverse probability weight that meets the \textit{balancing condition} introduced in Wallace and Moodie \cite{wallace2015doubly}, the dWOLS method leads to consistent estimators. The estimator is, therefore, called \textit{doubly-robust}. Note, a correct specification for one of the treatment or treatment-free models requires that i) the correct set of confounders be included in that model and ii) the confounders be incorporated as predictors in the model using the appropriate functional form (e.g., a variable that is deemed to affect quadratically the outcome must be included as a quadratic term in the outcome mean model). As mentioned in their introductory paper \cite{wallace2015doubly}, the well known inverse probability of treatment (IPT) weights \cite{horvitz1952generalization,rosenbaum1987model} meet the balancing condition. 

Wallace and Moodie\cite{wallace2015doubly} assumed that the observation times were not driven by covariates. In the current work, the data generating mechanism assumed for each time $t$ and for each individual $i$ is depicted in Figure \ref{fig2}. In that causal diagram,  the set of covariates $\mathbf{V}_i(t)=\left\{ \mathbf{K}_i(t), A_i(t), \mathbf{Z}_i(t) \right\}$ affects the observation indicator at each time $t$. We use an \textmyfont{IIV} weight proposed by Lin et al. \cite{lin2004analysis} to create a pseudo-population \cite{hernan2006estimating} in which covariates are unassociated with the observation process. Under assumption (\ref{ratee2}), an \textmyfont{IIV} weight of the form
\begin{align*}
\rho_i(t; \bm{\gamma}) = \left[ \xi_i(t)\exp\left\{ \bm{\gamma}' \mathbf{V}_i(t) \right\} d\Lambda_0(t) \right]^{-1}
\end{align*}
can be used to break the association between covariates in the set $\mathbf{V}_i(t)$ and the observation indicator. The causal diagram depicted in Figure \ref{fig2:right} is assumed to represent the updated data generating mechanism after \textmyfont{IIV}-weighting. If the time axis used in the recurrent events model for observation indicators is the time since cohort entry,  $d\Lambda_0(t)$ cancels out across individuals at time $t$ and it need not to be estimated.\cite{buuvzkova2009semiparametric} Therefore, our estimated \textmyfont{IIV} weight is given by
\begin{align}
 \rho_i(t; \hat{\bm{\gamma}}) = \left[\xi_i(t)\exp\left\{ \hat{\bm{\gamma}}' \mathbf{V}_i(t) \right\}  \right]_,^{-1} \label{iivww}
\end{align}
where $\hat{\bm{\gamma}}$ are obtained from the Andersen and Gill model \cite{andersen1982cox}, a model that can be fit using standard software, e.g.~\textit{coxph} of the \texttt{survival} package in R \cite{survival-package}.\\
We fit the propensity score  \cite{rosenbaum1983central} using a logistic regression model and use the resulting IPT weight as our balancing weight in the dWOLS. In the logistic regression model, we include all potential confounders as predictors of the treatment $A_i(t)$. The IPT weight is given by
\begin{align*}
 w_i(t; \hat{\bm{\kappa}}) = \frac{ P(A_i(t)=1) }{  P(A_i(t)=1| \mathbf{K}_i(t); \hat{\bm{\kappa}} )} + \frac{ P(A_i(t)=0) }{ P(A_i(t)=0| \mathbf{K}_i(t); \hat{\bm{\kappa}})}
\end{align*}
at time $t$ for individual $i$ (with $\hat{\bm{\kappa}}$ the fitted parameters from the logistic regression model). If the predictors of the logistic regression model are not of the correct functional form, confounding may remain. However, one advantage of using dWOLS is that we may have a second chance at unbiasedness if we can correctly specify the functional form of the confounders in the outcome mean model thanks to double robustness.

  \begin{minipage}[t]{0.5\textwidth}
\centering
\begin{tikzpicture}[scale=0.50][
->,
shorten >=2pt,
>=stealth,
node distance=1cm,
pil/.style={
->,
thick,
shorten =2pt,}
]
 
 \node (1) at (0,0) {A$(t)$};
\node (2) at (5,-3) {$\mathbf{Z}(t)$};
\node(3) at (2.5, -7) {Y$(t)$};
\node (4) at (-2, -6) {dN$(t)$};
\node(5) at (-3.5, -2){$\mathbf{K}(t)$};
\node(6) at (7.5, -3){$\mathbf{Q}(t)$};

\draw[->,black] (2) to  (4);
 \draw[->,black] (1) to  (3);
 \draw[->,black] (5) to  (1);
 \draw[->,black] (2) to  (3);
 \draw[->,black] (5) to  (3);
 \draw[->,black] (1) to  (4);
 \draw[->,black] (5) to  (4);
 \draw[->,black] (6) to  (3);
 
  \draw[->,black] (1) to  (2);

\end{tikzpicture} 
\captionsetup{font=footnotesize}

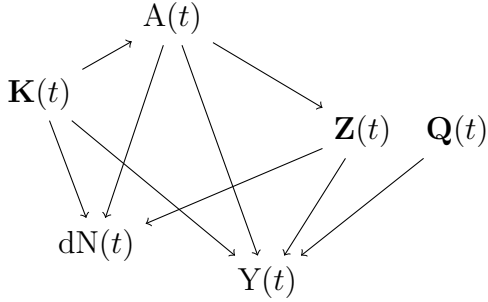
\captionof{figure}{Data generating mechanism considered in our simulation studies (the causal diagram is presented only for time $t$ and the individual index is removed). The set of covariates $\mathbf{V}(t)$ corresponds to $\left\{\mathbf{A}(t), \mathbf{Z}(t), \mathbf{K}(t) \right\}$ all affecting the observation indicator. Interactions are not depicted in the diagram.}\label{fig2}
\end{minipage}\hspace{.6cm}
\begin{minipage}[t]{0.5\textwidth}
 \centering
\begin{tikzpicture}[scale=0.5][
->,
shorten >=2pt,
>=stealth,
node distance=1cm,
pil/.style={
->,
thick,
shorten =2pt,}
]
 
 \node (1) at (0,0) {A$(t)$};
\node (2) at (5,-3) {$\mathbf{Z}(t)$};
\node(3) at (2.5, -7) {Y$(t)$};
\node (4) at (-2, -6) {dN$(t)$};
\node(5) at (-3.5, -2){$\mathbf{K}(t)$};
\node(6) at (7.5, -3){$\mathbf{Q}(t)$};
 
 \draw[->,black] (1) to  (3);
 \draw[->,black] (5) to  (1);
 \draw[->,black] (2) to  (3);
 \draw[->,black] (5) to  (3);
 \draw[->,black] (1) to  (2);
 
 \draw[->,black] (6) to  (3);
 
\end{tikzpicture} 
\captionsetup{font=footnotesize}

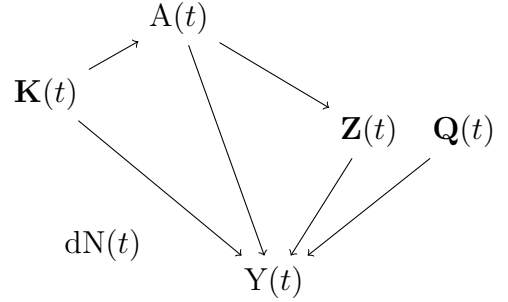
\captionof{figure}{Data generating mechanism after re-weighting data by the correctly specified \textmyfont{IIV} weight. Additional reweighting by IPT weights further removes the arrow from $\mathbf{K(t)}$ to $A(t)$. Interactions are not depicted in the diagram. }\label{fig2:right}
 
\end{minipage}\vspace{0.4cm}

  A weighted least squares (dWOLS in a one-stage treatment setting) that incorporates both weights is then fit to the data to estimate the blip function in (\ref{outcomee}). The proposed methodology corresponds to solving the following estimating equation for the coefficients of the mean outcome model in (\ref{outcomee})
  :\renewcommand{\arraystretch}{1}
\begin{align}
 &U(\bm{\beta},\bm{\psi}; \hat{\bm{\gamma}}, \hat{\bm{\kappa}} )= \sum_{i=1}^n \int_0^{\tau}w_i(t; \hat{\bm{\kappa}})\rho_i(t;\hat{\bm{\gamma}})\nonumber \\
& \hspace{2cm} \times\left[\begin{matrix}  \frac{\partial f\left\{ \mathbf{X}_i^{\beta}(t) ; \bm{\beta} \right\} }{\partial \bm{\beta}} \\  A_i(t)\mathbf{X}^{\psi}_i(t) \end{matrix} \right] \left[ Y_i(t)-  f\left\{ \mathbf{X}_i^{\beta}(t) ; \bm{\beta} \right\}  - A_i(t) \bm{\psi}'    \mathbf{X}^{\psi}_i(t)   \right] dN_i(t)= 0. \label{eepp}
\end{align}
\renewcommand{\arraystretch}{3}
In contrast to the estimator introduced in Coulombe et al.\cite{coulombe} for estimating marginal effects, the design matrix here includes not only interaction terms between the tailoring variables and treatment $A_i(t)$ but also the terms corresponding to potential confounders $\mathbf{K}_i(t)$ (leading to the double-robustness property, which was not addressed previously\cite{coulombe}). The proposed estimators for the parameters in the blip function are denoted by $\hat{\bm{\psi}}_{DW}$ (for \textit{Double Weights}) and those for the treatment-free model, by $\hat{\bm{\beta}}_{DW}$.  In simulation studies and in our application, we demonstrate empirically that our proposed methodology leads to a new type of robustness for the estimators in the blip function, which we term \textit{partially} doubly-robust. That is, in settings with confounding, only one of the propensity score model or outcome model must be correctly specified, the blip function must be correctly specified, and the observation model must only be correctly specified with respect to predictors that create an association between the treatment and the outcome. This mimics the coarseness of the propensity score, in that the observation model need not be the data generating model, but simply the coarsest function to provide balance with respect to patients' characteristics that also affect the longitudinal outcome, between instances with and without visits.
 In this work, we use a linear combination of the risk factors for the function $f\left\{\cdot\right\}$, i.e., we assume that $f\left\{ \mathbf{X}_i^{\beta}(t) ; \bm{\beta} \right\}= \bm{\beta}' \mathbf{X}_i^{\beta}(t)$ is an appropriate treatment-free model.  
 
 The asymptotic variance of the estimators $\hat{\bm{\beta}}_{DW}$ and $\hat{\bm{\psi}}_{DW}$ can be derived using theory on two-step estimators \cite{newey1994large} treating the parameters in the treatment and the observation models as nuisance parameters. Effectively, the asymptotic theory for the estimator for the optimal repeated measures ITR in equation (\ref{otr}) is obtained by reproducing the developments in Web Appendix C of Coulombe et al. \cite{coulombe} where the design matrix used to estimate the effect modifications is modified to include not only the treatment but also the tailoring variables, confounders, and pertinent interaction terms. In our application, we use nonparametric bootstrap with 500 samples to obtain variance estimates.

\section{Simulation study} \label{simul}

We conducted several simulation studies, with different strengths of dependence of the observation times on covariates, to assess the performance of our estimators $\hat{\bm{\psi}}_{DW}$. We used a data generating mechanism that was very similar to that presented in Figure \ref{fig2}. We compared our proposed estimator for the repeated measures ITR to three other types of estimator. The first, $\hat{\bm{\psi}}_{OLS}$ (for \textit{Ordinary Least Squares}), did not account for the observation process nor the confounders. This estimator consisted of estimating the blip function using dWOLS with no weights at all (but correctly modelling for the confounders in the treatment-free model). The second estimator, $\hat{\bm{\psi}}_{IPT}$, did not consider the covariate-driven observation process but incorporated an IPT weight as a function of a correctly specified propensity score. For the third strategy, we assessed our proposed estimator $\hat{\bm{\psi}}_{DW}$ under four different model misspecification scenarios. \textcolor{black}{The first scenario was based on all models specified correctly. The corresponding estimator is refered to as $\hat{\bm{\psi}}_{DW1}$ throughout the rest of the paper. We compared that scenario to i) a partially misspecified observation model and misspecified outcome model which lacked an adjustment for the second confounder $K_2$ ($\hat{\bm{\psi}}_{DW2}$); ii) a misspecified treatment model (that adjusted for the squared terms of $K_1$ and $K_3$ instead of their linear terms) and partially misspecified observation model ($\hat{\bm{\psi}}_{DW3}$); or iii) a (\textit{fully}) misspecified observation model ($\hat{\bm{\psi}}_{DW4}$). The partially misspecified observation model was such that the covariates required to appropriately block the biasing paths were correctly specified, but all other covariates were misspecified (more details follow in the next paragraph). Because of the partially double robustness of our estimator, the first three estimators are unbiased but may vary in their finite sample performance. In contrast, because of the misspecified observation model, the estimator $\hat{\bm{\psi}}_{DW4}$ has no guarantee of unbiasedness. We present in Supplementary Figure 1 (Supplementary Material A) the causal diagram corresponding to the data generating mechanism described below. Each estimator listed above incorporated different single or double weights which led to different pseudo-populations on which the mean outcome model was fitted. For each estimator, we present in Supplementary Figure 1 (panels b-g) the updated causal diagrams after the observations were reweighted by the corresponding weights and, based on the diagrams, provide there a justification for which estimators could be biased.}  

Our simulation studies were similar to those of Coulombe et al.\cite{coulombe} except that new tailoring variables were simulated and interaction terms were added in the outcome mean model. We tested sample sizes of either 250 or 500 patients and conducted 1000 simulations for each sample size. In the description that follows, we removed the patient index for ease of notation. First, three baseline confounders $\left\{K_{1},K_{2},K_{3}\right\}$ were generated with $K_{1}\sim \text{N}(1,1), K_{2}\sim \text{Bernoulli}(0.55)$, and $K_{3}\sim \text{N}(0,1)$. The treatment $A(t)$ was binary and time-varying and was simulated at each time $t$ as $A(t) \sim \text{Bernoulli}(p_{A})$ with $p_{A}= \text{expit} \left( 0.5 + 0.55\hspace{0.02cm} K_{1}-0.2\hspace{0.02cm} K_{2} -1 \hspace{0.02cm}K_{3}\right)$, where $\text{expit}(x)=\text{exp}\left\{x\right\}/\left\{1+\text{exp}(x)\right\} $. A time-varying mediator of the relationship between $A(t)$ and the outcome $Y(t)$ was simulated as a function of the treatment, as $Z(t)|A(t)=1 \sim \text{N}(2,1)$ and $Z(t)|A(t)=0 \sim \text{N}(4,2)$. A tailoring variable that did not depend on the treatment, the confounders, or the mediator was simulated as $Q(t)\sim\text{Bernoulli}(0.5)$. The outcome, before being set to \textit{missing} on times when the observation indicator equalled 0, was simulated for each time point as $Y(t)= \alpha(t) - 2\hspace{0.05cm} A(t) + 2.5\hspace{0.05cm} \left\{Z(t) - E\left[Z(t) |A(t)\right]\right\} +  0.4\hspace{0.05cm} K_{1}+ 0.05\hspace{0.05cm}K_{2} -0.6 \hspace{0.05cm} K_{3} + 0.5\hspace{0.05cm}\left\{A(t)\times Q(t) \right\}   -1\hspace{0.05cm}\left\{A(t)\times K_1\right\} +  \epsilon(t)$ with $ \epsilon(t) \sim \text{N}(\phi, 0.01)$, $\phi \sim \text{N}(0,0.04)$ an individual-specific random effect, and where the intercept function $\alpha(t)=\sqrt{t/100}$ was the same across individuals. Under this setup, the true values of the $\bm{\psi}$ parameters are (-2, 0.5, -1) and  correspond to the coefficients on the intercept, $Q(t)$ and $K_1$ in the blip function, respectively. \textcolor{black}{Note, the treatment effect is immediate in our simulation setting, as $A(t)$ is allowed to vary at any time and it affects $Y(t)$, an outcome measured roughly at the same time as $A(t)$ if we ignore the granularity of time discretization.} 

As discussed earlier, a challenge in the estimation of the optimal ITR is that the outcome is observed irregularly, and its observation depends on patients' covariates. To reproduce this behavior, the quantities above were first simulated in continuous time, with time discretized over a grid of 0.01 units, from 0 to $\tau=1$. Then, observation times (i.e.,~when the outcome was observed) were simulated according to a nonhomogeneous Poisson process, with intensity at time $t$ equal to $\lambda\left\{t| A(t),Z(t), K_2, K_3 \right\}=0.1 \hspace{0.05cm}\exp \left\{ \gamma_1 A(t)  + \gamma_2 Z(t)+\gamma_3 K_2 + \gamma_4 K_3 \right\}$. Bernoulli draws with probabilities proportional to these intensities were used at each time point to assign observation times (i.e., to determine whether the outcome is observed at that time). Observation times were drawn until the maximum follow-up time $\tau$. We tested different combinations of $\bm{\gamma}$ parameters, which encoded the strength of the dependence of observation indicators on covariates; these are shown in Table \ref{tab:char0}. \textcolor{black}{These parameters led to a different number of observation times of the outcome across individuals, which correspond to the repeated measures used for ITR estimation. Given that we simulated an immediate treatment effect, each observation (``repeated measure'') used for the ITR estimation comprised an observed outcome (e.g., $Y(s)$ at time $s$, for $dN(s)=1$) and its corresponding treatment $A(s)$ measured at the exact same time, along with confounders, tailoring variables and mediators.} To assess the performance of the proposed estimator when the observation model was misspecified, we estimated the observation intensity model using a) only $A(t)$ and $Z(t)$ as predictors (partial misspecification) or b) only $A(t)$ and $K_2$ as predictors (important misspecification). These strategies corresponded to estimators $\hat{\bm{\psi}}_{DW3}$ and $\hat{\bm{\psi}}_{DW4}$. Note, the latter model did not include the mediator $Z(t)$, a variable deemed to be important in the adjustment for the observation process because conditioning on $dN(t)$ opens a biasing path from $A(t)$ to $Y(t)$ that goes through $Z(t)$ and that is not due to the actual causal treatment effect. Furthermore, even if that observation model did contain an adjustment for $A(t)$ and $K_2$, the estimated coefficient for $A(t)$ in the observation model was likely biased, as $A(t)$ and $Z(t)$ were strongly dependent in the pseudo-population created after reweighting the observations by the IPT weights and the misspecified \textmyfont{IIV} weight in $\hat{\bm{\psi}}_{DW4}$. Hence, the observation model used in the estimator $\hat{\bm{\psi}}_{DW4}$ was likely misspecified with respect to both $A(t)$ and $Z(t)$, and the path going from $A(t)$ to $Y(t)$ via the mediator $Z(t)$ likely biased the causal effect of interest (see Supplementary Figure 1, panel e) for a depiction of the corresponding causal diagram).  
While the observation model in $\hat{\bm{\psi}}_{DW3}$ is misspecified, the estimator remains unbiased because the covariates that are not accounted for in the observation model are included in the IPT weights, thus blocking paths from the confounders into A(t). We therefore anticipated $\hat{\bm{\psi}}_{DW3}$ to be unbiased and possibly less variable than $\hat{\bm{\psi}}_{DW1}$ as the observation weights were expected to be more stable (as their models were more parsimonious).

In our simulation setting, the true value of the blip function (or ``gold standard'') \textcolor{black}{at time $t$} was given by $b( Q(t), K_1; \bm{\psi}=\left\{-2, 0.5,-1\right\} )= -2 + 0.5\hspace{0.05cm}Q(t) -1\hspace{0.05cm}K_1$.  We evaluated the performance of the estimators in three different ways. First, we computed the empirical mean squared error (MSE) of the blip values (i.e., the blip function evaluated at the covariates). For a given estimator $\hat{\bm{\psi}}=(\hat{\psi}_0, \hat{\psi}_1, \hat{\psi}_2 )$, that MSE was given by the mean of $\left[-2 + 0.5\hspace{0.05cm}q(t) -1\hspace{0.05cm}k_1 - ( \hat{\psi}_0 + \hat{\psi}_1 q(t) -\hat{\psi}_2 k_1) \right]^2$ (results averaged over 1000 simulations in Table \ref{tab:char0}).  
Next, we calculated the error rate in optimal treatment decisions (i.e., the proportion of estimated optimal treatment decisions that do not agree with the true optimal treatment decisions); see Supplementary Table 1 in Supplementary Material B. Our third performance criterion was the estimated value function, evaluated in a new population of size 25,000. In Supplementary Material B, we show the empirical bias of the six estimators for the blip values (i.e., the blip function evaluated at the observed covariates) (Supplementary Table 2), the absolute bias of each blip coefficient separately across all estimators compared and for each scenario for the observation process (Supplementary Table 3), and the performance in terms of value function (Supplementary Table 4).
\vspace{0.2cm}\\

\noindent \textit{Results of the simulation study}\\

The results of the comparison of MSE of the blip values across all six estimators for $\bm{\psi}$ (Table \ref{tab:char0}) are as expected. First, the MSE is larger (and similar) for $\hat{\bm{\psi}}_{DW4}$, $\hat{\bm{\psi}}_{OLS}$, and  $\hat{\bm{\psi}}_{IPT}$ (three last columns in Table \ref{tab:char0}) as compared with the three other estimators. Although $\hat{\bm{\psi}}_{DW4}$, $\hat{\bm{\psi}}_{OLS}$, and  $\hat{\bm{\psi}}_{IPT}$ make adjustments in the outcome mean model for confounders $\left\{K_1, K_2, K_3\right\}$, neither estimator accounts (adequately) for the observation process. The difference in MSE across the set of estimators $\hat{\bm{\psi}}_{DW4}$, $\hat{\bm{\psi}}_{OLS}$ and  $\hat{\bm{\psi}}_{IPT}$ and the set of estimators $\hat{\bm{\psi}}_{DW1}$, $\hat{\bm{\psi}}_{DW2}$, and $\hat{\bm{\psi}}_{DW3}$ increases with increasing sample size (Table \ref{tab:char0}). We obtain similar results when comparing the bias of the blip values, rather than the MSE (Supplementary Table 2). There is a clear decrease in the bias of the blip values for the estimators $\hat{\bm{\psi}}_{DW1}, \hat{\bm{\psi}}_{DW2}$, and  $\hat{\bm{\psi}}_{DW3}$ when increasing the sample size to 1000 or 2500, a result not observed with the three other estimators (Supplementary Table 2).

We observe similar patterns of results when comparing the treatment decisions from each estimator. Some scenarios for the parameters $\bm{\gamma}$, such as 2 and 3, leads to an important empirical bias in estimators $\hat{\bm{\psi}}_{DW4}$, $\hat{\bm{\psi}}_{OLS}$, and  $\hat{\bm{\psi}}_{IPT}$ which is reflected both in the MSE and empirical bias of the blip values and in the MSE of the optimal treatment decisions. Overall, the error rate of optimal treatment decisions varies from 0 to 6\% with the correctly specified estimators $\hat{\bm{\psi}}_{DW1}, \hat{\bm{\psi}}_{DW2}$, and  $\hat{\bm{\psi}}_{DW3}$, while it reaches 25\% with the three other estimators. For scenario 1, the error rate is small across all six estimators compared. This is explained by the fact that even when the blip function is biased, part of the treatment decisions is correct (unbiased) if the estimated blip value falls on the right side of the zero threshold (i.e., the threshold for the treatment rule).\renewcommand{\arraystretch}{1}
\begin{table}[H]
\caption{Simulation study results ($M=1000$ simulations) for the comparison of \textbf{MSEs} of the \textbf{blip values} obtained with six alternative models: DW1 the proposed doubly-weighted estimator which accounts for both processes correctly, DW2 for which the observation process was partially misspecified and the outcome model was misspecified, DW3 for which the treatment process was misspecified and the observation process was partially misspecified, DW4 for which the observation process was misspecified, OLS which does not adjust for confounding or observation process, and IPW which accounts only for confounding. Empirical MSEs are computed as the squared empirical bias of the estimated blip function evaluated at the patients' characteristics plus its empirical variance. The observation process varies but the confounding mechanism and the parameters of the true blip function remain the same in all 4 scenarios of varying $\bm{\gamma}$ below.}
\begin{center}
\begin{tabular}{ c c c  c  c c c c c }
 \hline
Sample& $\bm{\gamma}^{\upsilon}$   &  No.~obs. times &  \multicolumn{6}{c}{MSE}\\
size& parameters& mean (IQR)   & $\hat{\bm{\psi}}_{DW1}$ & $\hat{\bm{\psi}}_{DW2}$ & $\hat{\bm{\psi}}_{DW3}$ & $\hat{\bm{\psi}}_{DW4}$  &  $\hat{\bm{\psi}}_{OLS}$ &  $\hat{\bm{\psi}}_{IPT}$\\
      \hline
      250&1 &3 (1-3) & 0.60&0.46&0.32 & 0.84 & 0.75&0.83\\
                                        & 2 &3 (2-5) &1.69&1.86&1.60&3.24&2.88&3.26 \\
                              &       3 &6 (3-9) &1.99 &1.25&1.14&4.54&4.42&4.52\\
                               &        4 & 10 (8-12) &0.11&0.11&0.08&0.11&0.07&0.11\\ \hline

 500  &  1 &3 (1-3) &0.34 & 0.24&0.16&0.64&0.63&0.63 \\
                                         & 2 &3 (1-5) &0.92&1.06&0.84&2.82&2.61&2.83 \\
                 &                    3 &  6 (3-9) &1.27&0.72&0.66&4.36&4.30&4.36 \\ 
                    &    4         & 10 (8-12) &0.05&0.05&0.04 &0.05&0.03 &0.05 \\\hline

\end{tabular}
\end{center}
 \label{tab:char0}
 
 \noindent \scriptsize{$\upsilon$.1. (-2, -0.3, 0.2, -1.2);  
 2. (0.3, -0.6, -0.4, -0.3); 3. (0.4, -0.8, 1, 0.6);  4.  (0, 0, 0, 0), i.e., uninformative observation.\\
Abbreviations: MSE, mean squared error; IQR, interquartile range.}
\end{table}
 \renewcommand{\arraystretch}{3}
  In scenarios 2 and 3  for the $\bm{\gamma}$ parameters, the bias of the three ITR estimators using misspecified models is, therefore, due to biased blip values falling possibly near the threshold, yet on the wrong side of the threshold as compared to the true blip values.

The results above on the MSE and empirical bias of the blip values and error rates on the treatment decisions also agree with the results on the absolute bias of each blip coefficient found in Supplementary Table 3 (Supplementary Material B). In general, the bias in the estimation of the intercept coefficient of the blip function is small for the three preferred estimators $\hat{\bm{\psi}}_{DW1}, \hat{\bm{\psi}}_{DW2}$, and  $\hat{\bm{\psi}}_{DW3}$ and the bias decreases with the sample size. On the other hand, the intercept coefficient is consistently biased when estimated with $\hat{\bm{\psi}}_{DW4}$, $\hat{\bm{\psi}}_{OLS}$, or  $\hat{\bm{\psi}}_{IPT}$, even with increasing the sample size (Supplementary Table 3).

The performance as measured by the value function is also consistent with the other results above, showing a more important difference across the estimators under scenarios 2 and 3 for the observation process (Supplementary Table 4 in Supplementary Material B). The results for scenarios 2 and 3 convey the differences in optimal treatment decisions found across the correctly specified and the other estimators. Again, the results for the average outcome are similar across the three estimators $\hat{\bm{\psi}}_{DW1}$, $\hat{\bm{\psi}}_{DW2}$ and $\hat{\bm{\psi}}_{DW3}$, always leading to larger (or equal) average outcomes when compared to $\hat{\bm{\psi}}_{DW4}$, $\hat{\bm{\psi}}_{OLS}$, or  $\hat{\bm{\psi}}_{IPT}$. 
 
The results in this section and in Supplementary Material B demonstrate empirically that $\hat{\bm{\psi}}_{DW1}$, $\hat{\bm{\psi}}_{DW2}$ and $\hat{\bm{\psi}}_{DW3}$ lead to comparable MSEs of the blip values and error rates of the optimal treatment decisions, which MSEs and error rates are smaller when compared to more naive estimators that do not account appropriately for both the treatment and the observation processes. The proposed estimator $\hat{\bm{\psi}}_{DW}$ is unbiased for the optimal ITR under certain assumptions for the treatment, the observation, and the outcome models. We introduced the term \textit{partially doubly-robust} for our proposed methodology, where our treatment, observation, and outcome models have different opportunities to yield consistent estimators. If the observation model and the blip model are correctly specified, then only one of the treatment or the treatment-free models must be correctly specified. This is also true when the observation model is partially misspecified, that is, when it is misspecified only with respect to covariates that are not linking the treatment and the outcome in the causal diagram (assuming that the diagram also accounts for the paths blocked by the IPT weighting, if an IPT weight is used simultaneously). When the observation model is misspecified with respect to covariates linking the treatment and the outcome in the causal diagram after conditioning on the observation indicator $dN(t)$ and after re-weighting observations by a correctly specified IPT weight, then our proposed methodology generally fails to be consistent (depending on the strength of dependence between covariates and observation indicators).

\section{\textcolor{black}{Illustration} with the CPRD}

\begin{figure}[H]
\begin{center}

\begin{tikzpicture}[%
->,
shorten >=1pt,
>=stealth,
node distance=0.5cm,
pil/.style={
->,
thick,
shorten =1pt,}
]

\node (1) at (0,0) {Antidepressant$(t)$ (C/F) };
 
\node(3) at (2.5, -6) {BMI change utility$(t)$};
\node (4) at (-2, -6) {dN$(t)$};
\node(5) at (-3.5, -2){Confounders$(t)$ \footnotetext{Age, sex, age$*$sex, smoking, alcohol abuse, calendar year of cohort entry, psychiatric disease history, any other psychotropic drug prescriptions, lipid-lowering drugs, number of psychiatric admission in 6 months prior to entry, Index of Multiple Deprivation. } };
 
 \draw[->,black] (1) to  (3);
 \draw[->,black] (5) to  (1);
 
 \draw[->,black] (5) to  (3);
 \draw[->,black] (1) to  (4);
 \draw[->,black] (5) to  (4);

\end{tikzpicture}
\end{center}
\caption{Data generating mechanism at time $t$ considered in the application to CPRD. C/F respectively refer to citalopram and fluoxetine.}\label{fig3}
\end{figure}
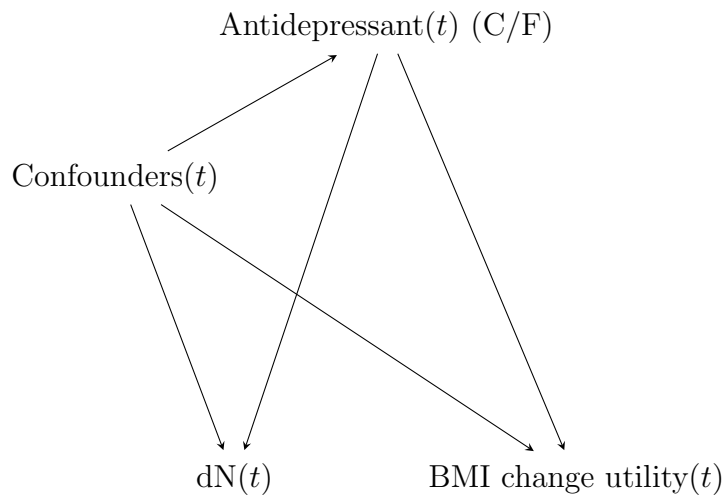

We applied the proposed methodology for the estimation of a repeated measures ITR to data from the UK's CPRD. Our aim was to develop an optimal ITR that chooses between two commonly prescribed antidepressant drugs, citalopram and fluoxetine, to optimize a BMI utility function. \textcolor{black}{The BMI utility function was repeatedly and irregularly measured in time.} We assumed that confounding and covariate-driven observation times were potential concerns and could cause bias in the estimation of the ITR. The causal diagram we assumed at each time is depicted in Figure \ref{fig3}. One potential source of bias due to the covariate-driven observation process is due to the (blocked) path going from the node \textit{Antidepressant}$(t)$ to $dN(t)$ and then to \textit{Confounders}$(t)$ and to the \textit{BMI change utility}$(t)$ in Figure \ref{fig3}. Upon conditioning on $dN(t)$, that path opens at the node $dN(t)$ and is unblocked from the treatment of interest to the BMI outcome. Another path that may bias the optimal ITR estimator is the backdoor path linking the nodes \textit{Antidepressant}$(t)$ to \textit{Confounders}$(t)$ and then to \textit{BMI change utility}$(t)$. The bias due to that backdoor path is called confounding.\\ 

\noindent \textit{Data source} \vspace{0.2cm}\\
\noindent CPRD is one of the largest primary care databases of anonymized health records and comes from a network of more than 700 general practictioner practices in the UK. The data contain demographics, lifestyle factors, prescription drugs, medical diagnoses, and referrals to specialists and hospitals for more than 13 million patients. \textcolor{black}{Information on prescription drugs comes from written prescriptions (as opposed to filled medications).} The data we used were linked with the Hospital Episode Statistics, which contains information on hospital diagnoses, and the Office for National Statistics mortality database, which provides details on dates and causes of death. The study protocol was approved by the Independent Scientific Advisory Committee of the United Kingdom Clinical Practice Research Datalink (CPRD) (protocol number 19\_017R) and the Research Ethics Committee of the Jewish General Hospital (Montr\'eal, Qu\'ebec, Canada).

\renewcommand{\arraystretch}{1}
\renewcommand{\arraystretch}{3}
We defined a cohort of new users of citalopram or fluoxetine with a recent diagnosis of depression. That cohort was previously defined and described\cite{coulombe2021can} and a flow chart for the cohort creation is shown in Supplementary Material C. Briefly, patient follow-up started at the time of initiation of either citalopram or fluoxetine, and follow-up was stopped (censored) when a patient discontinued their treatment, switched to another antidepressant drug, became pregnant, died, reached the end of registration with the practice, or the end of the study period (December 2017), whichever occurred first. \textcolor{black}{Although we censored patients' follow-up time when they discontinued or switched treatment, we were interested in a repeated measures ITR, in which treatment decisions are taken not only at cohort entry, but also at anytime during follow-up when a treatment decision must be made to reduce the potential weight change.} We only kept in the study cohort patients who had at least one BMI measurement before or at cohort entry and used the most recent BMI measurement to define their baseline BMI. Then, any BMI measurement recorded during patient follow-up was kept as an outcome for analysis (\textcolor{black}{i.e., an outcome for which we aim to optimize the expectation under the ITR}). If a BMI value was smaller than 15 or larger than 50, it was replaced by a missing value and not used in the analysis. \\

\noindent \textit{Outcome definition} \vspace{0.2cm}\\
\noindent For the \textcolor{black}{repeated} outcome, we defined a continuous utility function that conveyed the negative impacts of weight gain or weight loss while being treated with antidepressant drugs. \textcolor{black}{That outcome was defined every time when BMI was measured} as:
$$ U(t)= 100 -  5\times \mathbb{I}[\hspace{0.05cm}\text{Detrimental change in BMI$(t)$ category}\hspace{0.05cm} ] $$
$$ + \hspace{0.1cm}\mathbb{I}[ \hspace{0.05cm} \text{BMI}(0)<18.5\hspace{0.1cm} \cup \hspace{0.1cm} ( 18.5 \leq \text{BMI}(0) \leq 24.9 \hspace{0.1cm} \cap \hspace{0.1cm} \text{BMI}(t)<20) \hspace{0.05cm} ] \times \left\{\text{\% increase BMI$(t)$} \right\}$$
$$ -\hspace{0.1cm} \mathbb{I}[ \hspace{0.05cm}\text{BMI}(0) \ge 25   \hspace{0.1cm} \cup \hspace{0.1cm} \text{($18.5 \leq $ BMI$(0)\leq24.9$}  \hspace{0.1cm}\cap  \hspace{0.1cm} \text{BMI$(t)> $ 23.5)} \hspace{0.05cm} ] \times \left\{ \text{\% increase BMI$(t)$} \right\},$$ 

\noindent where the BMI categories are given by 
 \begin{align*}
 & <18.5 \hspace{0.2cm} & \text{Underweight} \\
 & 18.5 - 24.9 \hspace{0.2cm} & \text{Normal weight} \\
 & 25-29.9 \hspace{0.2cm} & \text{Overweight} \\
 & \ge 30 \hspace{0.2cm} & \text{Obese}
 \end{align*}
 \noindent and where the symbols $\cup$ and $\cap$ are respectively the logical ``or'' and ``and'' (such that either one or both conditions must be met). The ``\% increase BMI$(t)$'' is the relative change in BMI at time $t$ as compared to BMI measured at cohort entry. The coefficient of 5 in front of the first indicator function has been chosen arbitrarily to yield a reasonable shift in the utility distribution when a detrimental BMI change occurs. We considered as a ``detrimental change in BMI category'' a change from any category to another category other than normal with the exception of a change from obese to overweight. Remaining within the same category (e.g., within the overweight range) is not considered as a detrimental change in BMI category, but the utility function accounts for whether or not someone who is over weight loses weight to fall into normal BMI range (beneficial), gains weight (detrimental), or loses too much weight and becomes underweight (detrimental);
or whether an underweight patient loses weight (detrimental), gains too  much weight (detrimental), or gains just enough weight to fall in the normal BMI range, using the $\%$ increase in BMI from the utility formula. To give some latitude, we set the formula such that someone starting the study in the \textit{Normal weight} category has a utility function of 100 at time $t$ if their BMI is within the range [20, 23.5] at that time. The utility function's highest value corresponds to the best weight change outcome (either a beneficial weight change, if $U(t)>100$, or the absence of a detrimental weight change, if $U(t)=100$). As an individual's change in BMI varies in a detrimental fashion, the utility decreases towards some minimum across the study cohort. The minimum utility observed in our analysis dataset is reported in the Results section. \textcolor{black}{Theoretically, assuming that a patient may not have a percent change in BMI greater than 50\% since cohort entry (whether it is a detrimental or a beneficial change), the minimum value the utility can achieve is 45\%, and the maximum is 150\%. Given our outcome definition, a larger value for the utility function is better.} \vspace{0.1cm}\\
 
\noindent \textit{Confounder, tailoring variables, and predictors of observation} \vspace{0.2cm}\\
\noindent We defined the potential confounders of the relationship between the prescribed antidepressant and the utility function at baseline (cohort entry) and included the continuous-valued age, sex, current smoking status (smoker or non-smoker), alcohol abuse, calendar year of cohort entry (1998-2005, 2006-2011, 2012-2017), psychiatric disease history (which included autism spectrum disorder, obsessive-compulsive disorder, bipolar disorder, and schizophrenia), anxiety or generalized anxiety disorder (further referred to as anxiety), antipsychotics use, any other psychotropic medication use (benzodiazepine drugs, anxiolytics, barbiturates and hypnotics), lipid-lowering drugs, the number of psychiatric admissions or hospitalisations for self-harm in the 6 months prior to cohort entry and the Index of Multiple Deprivation \cite{deas2003measuring} as a proxy for the socioeconomic status. For the tailoring variables used to construct the optimal repeated measures ITR, similar to previous work \cite{coulombe2021can}, we included in the models the interaction terms between the treatment and age, sex, smoking status, a composite indicator of psychiatric disease history (a diagnosis for either autism spectrum disorder, obsessive-compulsive disorder, bipolar disorder, or schizophrenia), a diagnosis for anxiety, and the number of psychiatric admissions or hospitalisations for self-harm in the previous 6 months. \textcolor{black}{Note, we could also have defined time-varying confounders and tailoring variables, which our methodology allows for, but we used simpler definitions for our illustration.} Comorbidities were defined using any diagnostic codes recorded by cohort entry, and medication use, using any prescriptions in the year prior to cohort entry. 

In the observation intensity model for the outcome, we included the same set of covariates as confounders but these were defined in a time-varying manner. \textcolor{black}{Our rationale for this is that we wish to capture any effect of these variables on the observation intensity, and time-varying variables may, therefore, provide more sensitivity to any such effect}. For those time-varying covariates, we used different definitions. We considered a patient exposed to a medication for the duration of the corresponding prescription (medication considered were the lipid-lowering drugs, antipsychotics, and other psychotropic drugs). Then, after any first diagnosis for a chronic disease (including alcohol abuse, anxiety, and other psychiatric diseases), a patient was considered to have the condition for the remainder of the follow-up. The smoking status was updated at any time a new code related to smoking was recorded during follow-up (this included codes for smoking status and smoking cessation therapy). At any other time, it was defined using the most recent code for smoking. 

\textcolor{black}{As with the rest of the manuscript, we assumed in the illustration that the treatment was known (observed) at all times, which is realistic given that we had access to any prescriptions given by general practitioners. We also assumed that covariates in the two nuisance models (observation and treatment models) were always measurable.} However, the smoking status at baseline and the Index of Multiple Deprivation were missing for some individuals. We assumed that individuals with no smoking status recorded (at anytime before cohort entry) were non-smokers, and we removed from the study the few individuals ($<$1\%) with a missing Index of Multiple Deprivation. For all other covariates (confounders, tailoring variables or time-dependent features in the observation intensity model), we assumed that any existing condition was recorded in the database and that any drug prescribed was also recorded and available for defining the medication variables (and therefore, had no missing values).

  In the final outcome model, the design matrix incorporated the potential confounders, the treatment (citalopram or fluoxetine), and the interaction terms corresponding to all tailoring variables. The model, which predicted the utility function defined above, also incorporated both the IPT weight as a function of the propensity score and the \textmyfont{IIV} weight computed using the Andersen and Gill model. All predictors were included as linear terms in both nuisance models. Only the times when the utility function was available (i.e., when BMI was available) were included in the analysis and accounted for in the model fit. Those times corresponded to the repeated measures in the repeated measures ITR. We compared four different estimators for the optimal repeated measures ITR, with three of which were defined in Section \ref{simul}. The other estimator, $\hat{\bm{\psi}}_{IIV}$, is an \textmyfont{IIV}-weighted one-stage dWOLS estimator that incorporates an \textmyfont{IIV} weight. For each estimator and corresponding coefficients in the rule, we computed 95\% bootstrap confidence intervals (CIs) using 500 bootstrap samples. The bootstrap procedure considered the within-patient correlation by using a two-stage sampling where we first sampled patients with replacement (using the same sample size as the original dataset) and, within each patient, sampled the same number of measurements as the original dataset with replacement. The same procedure was used to obtain 95\% CIs for the observation rate ratios.\vspace{0.1cm}\\
 
\noindent \textit{Results} \vspace{0.2cm}\\
\textcolor{black}{After applying the exclusion criteria, the final cohort comprised 31,120 patients (60\% citalopram initiators) with a total of 48,388 records for BMI during follow-up that were used for estimating the repeated measured ITR.}  A comparison of patients' covariates at cohort entry, stratified by the study antidepressant, is presented in Supplementary Table 5 (Supplementary Material D). Some differences were found across the two groups, especially for the distribution of calendar year at cohort entry and the proportion of patients diagnosed with anxiety. These variables may act as confounders for the relationship between the antidepressants and the weight utility function.

  The estimated rate ratios obtained from the observation model are presented in Supplementary Table 6 (Supplementary Material D) along with their 95\% bootstrap CIs. A few variables were found to be associated with the observation intensity. Prescription for citalopram, male sex, and later calendar year of cohort entry were statistically significantly associated with lower chances for the weight to be recorded. On the other hand, a higher Index of Multiple Deprivation quintile, being an ever smoker, or the use of antipsychotics, other psychotropic drugs, or lipid-lowering drugs, were all statistically significantly associated with it being more likely for the outcome to be recorded.

 \textcolor{black}{When aggregating the BMI records, we found a crude mean utility function of 98.1 (range 46-148; SD 9.3) in patients who initiated citalopram and of 98.4 (range 46-146; SD 9.2) in those who initiated fluoxetine}. We present in Supplementary Table 7 (Supplementary Material D) the fitted coefficients for the tailoring variables in the mean outcome model, along with the corresponding 95\% bootstrap CIs. The fitted optimal ITR under our proposed methodology is given by: \vspace{0.2cm}\\
Treat with citalopram if
$$ -1.45 +0.16\times\mathbb{I}\left[\text{Male sex} \right]+0.13\times[\text{Index of Multiple Deprivation}]+0.08\times\mathbb{I}\left[ \text{Ever smoker} \right] $$
$$ +0.42\times\mathbb{I}\left[ \text{Alcohol abuse} \right]+1.31\times\mathbb{I}\left[ \text{Psychiatric diagnosis}\right]+0.35\times\mathbb{I}\left[ \text{Anxiety}\right]$$
$$-0.91\times\mathbb{I}\left[ \text{Antipsychotics drug use} \right] +0.30\times\mathbb{I}\left[  \text{Other psychotropic drug use}\right]$$
$$+0.21 \times\mathbb{I}\left[  \text{Lipid lowering drug use}\right]>0, $$
where $ \mathbb{I}\left[\cdot \right]$ is the indicator function. Age is not included as a tailoring variable in the rule, as its coefficient was null. \textcolor{black}{To provide an idea of the blip values one could obtain with this rule, we evaluated the rule under some of the 1280 possible profiles of patient characteristics; these results are shown in Supplementary Material E.} Finally, a comparison of the average fitted outcome under all ITR estimators is presented in Table \ref{tab:fitt}. Each outcome is fitted using the corresponding model applied to the actual treatment received (first row) or to the optimal treatment decision based on the rule (second row or Table \ref{tab:fitt}). Using optimal treatment rules to make the optimal treatment decision consistently leads to greater average fitted outcomes than the actual treatment received, and all four treatment rules lead to similar results in terms of optimization, in this case. In all four cases, we gain between 0.7 and 0.8\% utility when using an optimal treatment based on the estimated optimal ITR at times when the utility function is observed. The differences across the four estimated treatment rules are relatively modest in this study where the observation process was not very strongly linked to covariates (as per the rate ratios found in Supplementary Table 6) and where there were relatively few imbalances between treatment groups at cohort entry, such that confounding is relatively minor (Supplementary Table 5).
\renewcommand{\arraystretch}{1}
\begin{table}[H]
 \begin{center}
\caption{Comparison of the fitted outcome (i.e., a BMI-related utility function) under each estimated optimal ITR and compared to the actual treatment received, comparison for each of the four estimators: OLS which does not adjust for confounding or observation process, IPW which accounts only for confounding, \textmyfont{IIV} which accounts only for the observation process, and the proposed doubly-weighted estimator which accounts for both processes, CPRD, UK, 1998-2017, $n=31,120$ individuals. }
\begin{tabular}{ l c c c c  }
 \hline  \hline
  &  \multicolumn{4}{c}{Average fitted outcome (SE$^\dagger$)}     \\
 Treatment & $\hat{\bm{\psi}}_{OLS}$   & $\hat{\bm{\psi}}_{IPT}$& $\hat{\bm{\psi}}_{IIV}$ &   $\hat{\bm{\psi}}_{DW1}$    \\ \hline
Actual treatment received & 98.2 (0.001)&98.2 (0.001) &98.3 (0.001) &98.3 (0.001) \\
Optimal treatment & 98.9 (0.001) &98.9 (0.001) &99.1 (0.001) &  99.0 (0.001) \\
 \hline
\end{tabular}
 \label{tab:fitt}
  
  \scriptsize{ Abbreviation: SE, standard error. \\
  $\dagger$. Based on prediction SEs obtained from the model that were further summed and normalized to obtain the variance of the mean of all predicted outcomes, rather than the variance of individual predicted outcome values. 
  }
   \end{center}

 \end{table}
 \renewcommand{\arraystretch}{3}
  One difference in the four optimal treatment rules is found in the coefficient of the interaction term between the treatment and diagnosis for anxiety, where the doubly-weighted proposed estimator is the only approach leading to an effect modification that is statistically significant at the 0.05 level (Supplementary Table 7). It is a signal that patients' anxiety may generally be useful in tailoring the antidepressant drug, after accounting for the covariate-driven treatment and observation processes. To generalize these results, however, the study should be reproduced in other (possibly larger) study cohorts.

\section{Discussion}
 
 In observational studies using longitudinal data extracted from EHR, patients are often observed at irregular times that may depend on their own characteristics. When these same characteristics are associated with the treatment and/or the outcome, causal inference on treatment effects can be affected. In developing optimal ITRs that rely directly on those treatment effects, it is important to determine how observation times can impact the inference. Drawing causal diagrams \cite{pearl1998graphs, hernan2009observation} can help in finding potential biasing paths between the treatment and the outcome that should be blocked via, e.g., IPT weighting or \textmyfont{IIV} weighting.
 
 In this work, we proposed a novel methodology to account for covariate-driven treatment and observation mechanisms simultaneously in the estimation of optimal repeated measured ITRs. In extensive simulation studies, we demonstrated the consistency of the methodology. The proposed method is a straightforward extension of previous work \cite{coulombe}, and the same asymptotic theory can be used to develop the asymptotic variance of our proposed estimators. Our method is easy to implement and more easily understood than methods such as g-estimation. We applied the method to data from the UK's CPRD and proposed an optimal ITR for choosing between citalopram and fluoxetine (two commonly prescribed antidepressants) to treat depression while reducing BMI changes that could be detrimental for one's health.
 
 The proposed methodology relies on assumptions that are commonly postulated in the literature on causal inference. First, our adjustment sets for both the treatment and the observation models should contain enough covariates so as to break any biasing association between the treatment and outcome due to confounders or to observation indicators. Secondly, we postulated positivity of treatment and observation, which can be unrealistic in certain settings with EHR. However, coarsening of the data in time may be used to circumvent non positivity issues. The work of Robins et al.\cite{robins2008estimation} (Section 6) and Neugebauer et al. \cite{neugebauer2017identification} also allowed causal inference under a weaker positivity assumption for the monitoring and could possibly be extended to our setting. Moreover, given that our work focuses on one-stage ITRs (as opposed to multiple stages DTRs), the positivity assumption for treatment is weaker (i.e., easier to meet) than the one typically made when building multiple decisions rules where long sequences of treatment must have a non-zero chance of occurring. In this work, we did not consider a sequence of treatments but rather the cross-sectional impact of a binary treatment and \textcolor{black}{we allowed each patient to contribute multiple measurements, one for each outcome observed}. We further assumed consistency of the outcome, which encompasses that the treatment definition be clear and that there be no interaction between individuals (no \textit{spillover} in treatment effects). In our setting, the latter assumption is realistic as the antidepressant drug taken by one patient is unlikely to affect another patient's weight, and the CPRD data are collected over a large geographic area (such that patients are unlikely to interact with most other patients in the study cohort). The treatment definition is also clear and simple in our application to the CPRD, such that we are not worried about treatment variations that might affect the consistency of the potential outcome. Finally, our method requires the standard assumptions \textcolor{black}{for one-stage repeated measures ITRs, i.e., treatment effects should be acute and there should be no antagonistic or synergistic effect due to previous treatments affecting the current one \cite{larry}. Although it is not certain that these assumptions were met in our illustration using CPRD data, most subsequent weight measurements were taken far apart in time, reducing the chances of a carryover effect from a previous treatment.}
 
In future work, we aim to extend the proposed methodology to the more complex setting of the more traditional, multiple stage DTRs using dWOLS. In that setting, dWOLS has great advantages as it can incorporate weights that are cumulated over time (similarly to marginal structural models). As such, it is a method of choice for treating complex covariate-driven observation processes (such as those that depend on an endogenous covariate process\cite{coulombe2021estimation}) and time-dependent confounding, in which there can be a biasing feedback between the covariates and the processes.

\section*{Acknowledgements} 

The authors would like to thank Dr. Shannon Holloway for reading the manuscript and providing significant help with editing. This research was enabled in part by support provided by Compute Canada (www.computecanada.ca). Computations were performed on the Niagara supercomputer at the SciNet HPC Consortium. SciNet is funded by the Canada Foundation for Innovation; the Government of Ontario; Ontario Research Fund - Research Excellence; and the University of Toronto.

\section*{Funding}
Part of the research was funded by an Innovative Ideas Award from Healthy Brains for Healthy Lives [HBHL 1c-II-11]. Dr. Coulombe and Dr. Shortreed are supported by the National Institute of Mental Health of the National Institutes of Health [R01 MH114873]. Dr. Moodie holds a Canada Research Chair (Tier 1) in Statistical Methods for Precision Medicine. Dr. Moodie further acknowledges support from a Discovery Grant from the Natural Sciences and Engineering Research Council (NSERC) and a chercheur de m\'erite career award from the Fonds de recherche du Québec--Santé. Dr. Renoux is supported by a chercheur-boursier salary award from the Fonds de recherche du Québec-Santé.

\section*{Declaration of conflicting interests}
 Dr. Shortreed has worked on grants awarded to Kaiser Permanente Washington Health Research Institute (KPWHRI) by Bristol Meyers Squibb and by Pfizer. She was also a co-Investigator on grants awarded to KPWHRI from Syneos Health, who represented a consortium of pharmaceutical companies carrying out FDA-mandated studies regarding the safety of extended-release opioids. Authors JC, EEMM and CR have no conflict of interest to declare. The content is solely the responsibility of the authors and does not necessarily represent the official views of the National Institutes of Health.  
 
   \bibliographystyle{unsrtnat}

\begin{thebibliography}{9}
\bibitem{bang2005doubly}
Bang H and Robins JM. Doubly robust estimation in missing data and causal inference models. \textit{Biometrics} 2005; 61(4): 962--973.
 
 \bibitem{stuart2010matching}
 Stuart EA. Matching methods for causal inference: A review and a look forward. \textit{Stat Sci} 2010; 25(1): 1--21.
 
 
\bibitem{moodie2012q}
 Moodie EEM, Chakraborty B and Kramer MS. Q-learning for estimating optimal dynamic treatment rules from observational data. \textit{Can J Stat} 2012; 40(4): 629--645.
 
 \bibitem{schuler2017targeted}
Schuler MS and Rose S. Targeted maximum likelihood estimation for causal inference in observational studies. \textit{Am J Epidemiol} 2017; 185(1): 65--73.

 \bibitem{kosorok2019precision}
Kosorok MR and Laber EB. Precision medicine. \textit{Annu Rev Stat App} 2019; 6: 263--286.

 \bibitem{robins1992g}
Robins JM, Blevins D, Ritter G and Wulfsohn M. G-estimation of the effect of prophylaxis therapy for pneumocystis carinii pneumonia on the survival of AIDS patients. \textit{Epidemiology} 1992; 3(4): 319--336.

 \bibitem{laber2014q}
 Laber EB, Rose EJ, Davidian M and Tsiatis AA. Q-Learning. \textit{Wiley StatsRef: Statistics Reference Online} 2014: 1--10.
 
 \bibitem{wallace2015doubly}
 Wallace MP and Moodie EEM. Doubly-robust dynamic treatment regimen estimation via weighted least squares. \textit{Biometrics} 2015; 71(3): 636--644.
 
 
 \bibitem{larry}
Dong L, Moodie EEM, Villain L and Thiébaut R. Evaluating the use of generalized dynamic weighted ordinary least squares for individualized HIV treatment strategies. \textit{arXiv preprint} 2021; arXiv:2109.01218.
 
 \bibitem{wallace2017dynamic}
Wallace MP, Moodie EEM and Stephens DA. Dynamic treatment regimen estimation via regression-based techniques: Introducing R package DTRreg. \textit{J Stat Softw} 2017; 80(1): 1--20.
 
 \bibitem{wallace2016package}
 Wallace MP, Moodie EEM and Stephens DA. Package `DTRreg'. \emph{R package version 1.7}, 2016. URL: \url{https://cran.r-project.org/web/packages/DTRreg/index.html}.
 
 \bibitem{simoneau2020optimal}
 Simoneau G, Moodie EEM, Azoulay L and Platt RW. Adaptive treatment strategies with survival outcomes: An application to the treatment of type 2 diabetes using a large observational database. \textit{Am J Epidemiol} 2020; 189(5): 461--469.
 
 \bibitem{tsiatis2019dynamic} 
Tsiatis AA, Davidian M, Holloway ST and Laber EB. \textit{Dynamic treatment regimes: statistical methods for precision medicine}. Boca Raton: CRC Press, 2019. 
 
\bibitem{linn}
Linn KA, Laber EB and Stefanski LA. iq{L}earn: interactive {Q}-learning in {R}. \textit{J Stat Softw} 2015; 64(1): 1--32.

 \bibitem{mcgrath2020gformula}
McGrath S, Lin V, Zhang Z, Petito LC, Logan RW, Hern{\'a}n MA and Young JG. gfo{R}mula: an {R} package for estimating the effects of sustained treatment strategies via the parametric g-formula. \textit{Patterns} 2020; 1(3): 1--12. 
  
 \bibitem{horvitz1952generalization}
 Horvitz DG and Thompson DJ. A generalization of sampling without replacement from a finite universe. \textit{J Am Stat Assoc} 1952; 47(260): 663--685.
 
 \bibitem{rosenbaum1987model}
 Rosenbaum PR. Model-based direct adjustment. \textit{J Am Stat Assoc} 1987; 82(398): 387--394.
 
 
 \bibitem{coulombe2021can}
 Coulombe J, Moodie EEM, Shortreed SM and Renoux C. Can the risk of severe depression-related outcomes be reduced by tailoring the antidepressant therapy to patient characteristics? \textit{Am J Epidemiol} 2021; 190(7): 1210--1219.
 
 \bibitem{johnson2018causal}
 Johnson KW, Glicksberg BS, Hodos RA, Shameer K, and Dudley JT. Causal inference on electronic health records to assess blood pressure treatment targets: An application of the parametric g formula. \textit{Pacific Symposium on Biocomputing 2018: Proceedings of the Pacific Symposium} 2018: 180--191.
  
 
 \bibitem{greenland2003quantifying}
 Greenland S. Quantifying biases in causal models: Classical confounding vs collider-stratification bias. \textit{Epidemiology} 2003; 14(3): 300--306.
 
 
 
 \bibitem{coulombe2021estimation}
Coulombe J, Moodie EEM, Platt RW and Renoux C. Estimation of the marginal effect of antidepressants on body mass index under confounding and endogenous covariate-driven monitoring times. \textit{Ann Appl Stat} 2021; forthcoming.


 \bibitem{robins2008estimation}
Robins JM, Orellana L and Rotnitzky A. Estimation and extrapolation of optimal treatment and testing strategies. \textit{Stat Med} 2008; 27(23): 4678--4721.
 
  
 

 \bibitem{neugebauer2017identification}
Neugebauer R, Schmittdiel JA, Adams AS, Grant RW and van der Laan MJ. Identification of the joint effect of a dynamic treatment intervention and a stochastic monitoring intervention under the no direct effect assumption. \textit{J Causal Inference} 2017; 5(1): 1--66.
 
 \bibitem{kreif2018evaluation}
 Kreif N, Sofrygin O, Schmittdiel J, Adams A, Grant R, Zhu Z, van der Laan M and Neugebauer R. Evaluation of adaptive treatment strategies in an observational study where time-varying covariates are not monitored systematically. \textit{arXiv preprint} 2018; arXiv:1806.11153.
 
 \bibitem{thall2007bayesian}
 Thall PF, Wooten LH, Logothetis CJ, Millikan RE and Tannir NM. Bayesian and frequentist two-stage treatment strategies based on sequential failure times subject to interval censoring. \textit{Stat Med} 2007; 26(26): 4687--4702.
 
 \bibitem{shahn2019g}
 Shahn Z, Li Y, Sun Z, Mohan A, Sampaio C and Hu J. G-computation and hierarchical models for estimating multiple causal effects from observational disease registries with irregular visits. \textit{AMIA Jt Summits Transl Sci Proc} 2019: 789--798.
 
 \bibitem{yang2021semiparametric}
 Yang S. Semiparametric estimation of structural nested mean models with irregularly spaced longitudinal observations. \textit{Biometrics} in press. URL: \url{https://onlinelibrary.wiley.com/doi/10.1111/biom.13471}.   
 
 \bibitem{lok2017mimicking}
 Lok JJ. Mimicking counterfactual outcomes to estimate causal effects. \textit{Ann Stat} 2017; 45(2): 461-499.
 
 \bibitem{goldstein}
 Goldstein BA, Phelan M, Pagidipati NJ and Peskoe SB. How and when informative visit processes can bias inference when using electronic health records data for clinical research. \textit{J Am Med Inform Assoc} 2019; 26(12): 1609-1617.
 
 \bibitem{mcculloch}
 McCulloch CE, Neuhaus JM and Olin RL. Biased and unbiased estimation in longitudinal studies with informative visit processes. \textit{Biometrics} 2016; 72(4): 1315-1324.
 
 \bibitem{lin2004analysis}
 Lin H, Scharfstein DO and Rosenheck RA. Analysis of longitudinal data with irregular, outcome-dependent follow-up. \textit{J Roy Stat Soc B} 2004; 66(3): 791--813.
 
 \bibitem{buuvzkova2009semiparametric}
 Bůžková P and Lumley T. Semiparametric modeling of repeated measurements under outcome-dependent follow-up. \emph{Stat Med} 2009; 28(6): 987--1003.
 
 \bibitem{zhu2017estimation}
Zhu Y, Lawless JF and Cotton CA. Estimation of parametric failure time distributions based on interval-censored data with irregular dependent follow-up. \emph{Stat Med} 2017; 36(10): 1548--1567. 
 
 \bibitem{liang2009joint}
Liang Y, Lu W and Ying Z. Joint modeling and analysis of longitudinal data with informative observation times. \textit{Biometrics} 2009; 65(2): 377--384.
 
 \bibitem{cai2012time}
 Cai N, Lu W and Zhang HH. Time-varying latent effect model for longitudinal data with informative observation times. \emph{Biometrics} 2012; 68(4): 1093--1102.
 
 \bibitem{dai2018joint}
Dai H and Pan J. Joint modelling of survival and longitudinal data with informative observation times. \emph{Scand J Stat} 2018; 45(3): 571--589.
 
 \bibitem{lipsitz2002parameter}
 Lipsitz SR, Fitzmaurice GM, Ibrahim JG, Gelber R and Lipshultz S. Parameter estimation in longitudinal studies with outcome-dependent follow-up. \emph{Biometrics} 2002; 58(3): 621--630.
 
 \bibitem{coulombe}
 Coulombe J, Moodie EEM and Platt RW. Weighted regression analysis to correct for informative monitoring times and confounders in longitudinal studies. \emph{Biometrics} 2021; 77(1): 162--174.
 
 
 \bibitem{herrett2015data}
Herrett E, Gallagher AM, Bhaskaran K, Forbes H, Mathur R, Van Staa T and Smeeth L Data resource profile: {C}linical {P}ractice {R}esearch {D}atalink {(CPRD)}. \emph{Int J Epidemiol} 2015; 44(3): 827--836.

 \bibitem{neyman1923application}
 Neyman JS. On the application of probability theory to agricultural experiments. {E}ssay on principles. {S}ection 9 (translation published in 1990). \emph{Stat Sci} 1990; 5(4): 472--480.
 
 \bibitem{rubin1974estimating}
Rubin DB. Estimating causal effects of treatments in randomized and nonrandomized studies. \emph{J Educ Psychol} 1974; 66(5): 688--701.
 
 \bibitem{lawless1995some}
Lawless JF and Nadeau C. Some simple robust methods for the analysis of recurrent events. \emph{Technometrics} 1995; 37(2): 158--168.
 
 \bibitem{lin2001semiparametric}
Lin DY and Ying Z. Semiparametric and nonparametric regression analysis of longitudinal data. \emph{J Am Stat Assoc} 2001; 96(453): 103--126.
 
 \bibitem{hernan2006estimating}
 Hern{\'a}n MA and Robins JM. Estimating causal effects from epidemiological data. \emph{J Epidemiol Commun H} 2006; 60(7): 578--586.
 
 \bibitem{andersen1982cox}
 Andersen PK and Gill RD. Cox's regression model for counting processes: A large sample study. \emph{Ann Stat} 1982; 10(4): 1100-1120.
 
 \bibitem{survival-package}
 Therneau MT. A package for survival analysis in {R}. \textit{R package version 3.1-12}, 2018. URL: \url{https://CRAN.R-project.org/package=survival}.
 
 \bibitem{rosenbaum1983central}
 Rosenbaum PR and Rubin DB. The central role of the propensity score in observational studies for causal effects. \emph{Biometrika} 1983; 70(1): 41--55.
 
 \bibitem{newey1994large}
 Newey KW and McFadden D. Large sample estimation and hypothesis in WK Newey and D McFadden (eds) \textit{Handbook of Econometrics}  4th ed. Amsterdam: Elsevier, 1994, pp. 2112--2245.
 
 
  
\bibitem{deas2003measuring}
Deas I, Robson B, Wong C and Bradford M. Measuring neighbourhood deprivation: A critique of the Index of Multiple Deprivation. \emph{Environ Plann C} 2003; 21(6): 883--903.

 
 \bibitem{pearl1998graphs}
Pearl J. Graphs, causality, and structural equation models. \emph{Sociol Method Res} 1998; 27(2): 226--284.
 
 \bibitem{hernan2009observation}
Hern{\'a}n MA, McAdams M, McGrath N, Lanoy E and Costagliola D. Observation plans in longitudinal studies with time-varying treatments. \emph{Stat Methods Med Res} 2009; 18(1): 27--52.
  
 
\end{thebibliography}

\newpage
\appendix

\captionsetup[table]{name=Supplementary Table}
\captionsetup[figure]{name=Supplementary Figure}
\setcounter{table}{0}
\setcounter{figure}{0}
\begin{singlespace}

\renewcommand{\arraystretch}{1}

\noindent \textbf{Supplementary Material for ``Estimating Individualized Treatment Rules in Longitudinal Studies with Covariate-Driven Observation Times ''  }

\newpage
 
\noindent  \textbf{Supplementary Material A}\vspace{0.5cm}\\
  
\begin{figure} 

  \centering
  \begin{minipage}{.48\linewidth}
    \centering
    \subcaptionbox{}
      {\begin{tikzpicture}[scale=0.55 ][%
->,
shorten >=2pt,
>=stealth,
node distance=1cm,
pil/.style={
->,
thick,
shorten =2pt,}
]

 \node (1) at (0,0) {A$(t)$};
\node (2) at (2.5,-2) {$Z(t)$};
\node(3) at (1.8, -7) {Y$(t)$};
\node (4) at (0, -3.8) {dN$(t)$};
\node(5) at (-3.5, -1.){$K_1$};
\node(5b) at (-3.9, -2){$K_2$};
\node(5c) at (-3.5, -3){$K_3$};
\node(6) at (5.5, -3){$Q(t)$};

\draw[->,black] (2) to  (4);
 \draw[->,black] (1) to  (3);
 \draw[->,black] (5) to  (1);
  \draw[->,black] (5b) to  (1);
   \draw[->,black] (5c) to  (1);
 \draw[->,black] (2) to  (3);
 \draw[->,black] (5) to  (3);
  \draw[->,black] (5b) to  (3);
   \draw[->,black] (5c) to  (3);
 \draw[->,black] (1) to  (4);
 \draw[->,black] (5b) to  (4);
  \draw[->,black] (5c) to  (4);
 \draw[->,black] (6) to  (3);
 
  \draw[->,black] (1) to  (2);

\end{tikzpicture} }
 \vspace{0.1cm}
    \subcaptionbox{}
      {\begin{tikzpicture}[scale=0.55][%
->,
shorten >=2pt,
>=stealth,
node distance=1cm,
pil/.style={
->,
thick,
shorten =2pt,}
]

 \node (1) at (0,0) [draw]{A$(t)$};
\node (2) at (2.5,-2) {$Z(t)$};
\node(3) at (1.8, -7) {Y$(t)$};
\node (4) at (0, -3.8) [draw]{dN$(t)$};
\node(5) at (-3.5, -1.)[draw]{$K_1$};
\node(5b) at (-3.9, -2)[draw]{$K_2$};
\node(5c) at (-3.5, -3)[draw]{$K_3$};
\node(6) at (5.5, -3)[draw]{$Q(t)$};

 \draw[->,black] (1) to  (3);
 \draw[->,black] (2) to  (3);
 \draw[->,black, ] (5) to  (3);
  \draw[->,black, ] (5b) to  (3);
   \draw[->,black, ] (5c) to  (3);
 \draw[->,black] (6) to  (3);
 
  \draw[->,black] (1) to  (2);

\end{tikzpicture} }
 \vspace{0.1cm}
    \subcaptionbox{}
      {\begin{tikzpicture}[scale=0.55][%
->,
shorten >=2pt,
>=stealth,
node distance=1cm,
pil/.style={
->,
thick,
shorten =2pt,}
]

 \node (1) at (0,0)[draw] {A$(t)$};
\node (2) at (2.5,-2) {$Z(t)$};
\node(3) at (1.8, -7) {Y$(t)$};
\node (4) at (0, -3.8)[draw] {dN$(t)$};
\node(5) at (-3.5, -1.)[draw]{$K_1$};
\node(5b) at (-3.9, -2) {$K_2$};
\node(5c) at (-3.5, -3)[draw]{$K_3$};
\node(6) at (5.5, -3)[draw]{$Q(t)$};

 \draw[->,black] (1) to  (3);
 \draw[->,black] (2) to  (3);
 \draw[->,black, ] (5) to  (3);
  \draw[->,black, ] (5b) to  (3);
   \draw[->,black, ] (5c) to  (3);
 \draw[->,black] (5b) to  (4);
  \draw[->,black] (5c) to  (4);
 \draw[->,black] (6) to  (3);
 
  \draw[->,black] (1) to  (2);

\end{tikzpicture} }
      \vspace{0.1cm}
    \subcaptionbox{}
      {\begin{tikzpicture}[scale=0.55][%
->,
shorten >=2pt,
>=stealth,
node distance=1cm,
pil/.style={
->,
thick,
shorten =2pt,}
]

 \node (1) at (0,0)[draw] {A$(t)$};
\node (2) at (2.5,-2) {$Z(t)$};
\node(3) at (1.8, -7) {Y$(t)$};
\node (4) at (0, -3.8)[draw] {dN$(t)$};
\node(5) at (-3.5, -1.)[draw]{$K_1$};
\node(5b) at (-3.9, -2)[draw]{$K_2$};
\node(5c) at (-3.5, -3)[draw]{$K_3$};
\node(6) at (5.5, -3)[draw]{$Q(t)$};
   \draw[->,black, dashed] (1) to  (4);
 \draw[->,black] (1) to  (3);
 \draw[->,black] (5) to  (1);
   \draw[->,black] (5c) to  (1);
 \draw[->,black] (2) to  (3);
 \draw[->,black] (5) to  (3);
  \draw[->,black] (5b) to  (3);
   \draw[->,black] (5c) to  (3);
 \draw[->,black] (5b) to  (4);
  \draw[->,black] (5c) to  (4);
 \draw[->,black] (6) to  (3);
 
  \draw[->,black] (1) to  (2);

\end{tikzpicture} }

    \vspace{0.1cm}
  \end{minipage}\quad
  \begin{minipage}{.48\linewidth}
    \centering
    \subcaptionbox{}
      {\begin{tikzpicture}[scale=0.55][%
->,
shorten >=2pt,
>=stealth,
node distance=1cm,
pil/.style={
->,
thick,
shorten =2pt,}
]

 \node (1) at (0,0) [draw]{A$(t)$};
\node (2) at (2.5,-2) {$Z(t)$};
\node(3) at (1.8, -7) {Y$(t)$};
\node (4) at (0, -3.8)  [draw]{dN$(t)$};
\node(5) at (-3.5, -1.) [draw]{$K_1$};
\node(5b) at (-3.9, -2) [draw]{$K_2$};
\node(5c) at (-3.5, -3) [draw]{$K_3$};
\node(6) at (5.5, -3) [draw]{$Q(t)$};
\draw[->,black,dashed] (1) to  (4);
\draw[->,black] (2) to  (4);
 \draw[->,black] (1) to  (3);
 \draw[->,black] (2) to  (3);
 \draw[->,black] (5) to  (3);
  \draw[->,black] (5b) to  (3);
   \draw[->,black] (5c) to  (3);
  \draw[->,black] (5c) to  (4);
 \draw[->,black] (6) to  (3);
 
  \draw[->,black] (1) to  (2);

\end{tikzpicture} }
 \vspace{0.1cm}
    \subcaptionbox{}
      {\begin{tikzpicture}[scale=0.55][%
->,
shorten >=2pt,
>=stealth,
node distance=1cm,
pil/.style={
->,
thick,
shorten =2pt,}
]

 \node (1) at (0,0) [draw]{A$(t)$};
\node (2) at (2.5,-2) {$Z(t)$};
\node(3) at (1.5, -7) {Y$(t)$};
\node (4) at (0, -3.8) [draw]{dN$(t)$};
\node(5) at (-3.5, -1)[draw]{$K_1$};
\node(5b) at (-3.9, -2)[draw]{$K_2$};
\node(5c) at (-3.5, -3)[draw]{$K_3$};
\node(6) at (5.5, -3)[draw]{$\mathbf{Q}(t)$};
   \draw[->,black] (5) to  (3);
\draw[->,black] (2) to  (4);
 \draw[->,black] (1) to  (3);
 \draw[->,black] (2) to  (3);
 \draw[->,black] (5) to  (3);
  \draw[->,black] (5b) to  (3);
   \draw[->,black] (5c) to  (3);
 \draw[->,black] (1) to  (4);
 \draw[->,black] (5b) to  (4);
  \draw[->,black] (5c) to  (4);
 \draw[->,black] (6) to  (3);
 
  \draw[->,black] (1) to  (2);
 
\end{tikzpicture} }
 \vspace{0.1cm}
    \subcaptionbox{}
      {\begin{tikzpicture}[scale=0.55][%
->,
shorten >=2pt,
>=stealth,
node distance=1cm,
pil/.style={
->,
thick,
shorten =2pt,}
]

 \node (1) at (0,0) [draw]{A$(t)$};
\node (2) at (2.5,-2) {$Z(t)$};
\node(3) at (1.5, -7) {Y$(t)$};
\node (4) at (0, -3.8) [draw]{dN$(t)$};
\node(5) at (-3.5, -1)[draw]{$K_1$};
\node(5b) at (-3.9, -2)[draw]{$K_2$};
\node(5c) at (-3.5, -3)[draw]{$K_3$};
\node(6) at (5.5, -3)[draw]{$\mathbf{Q}(t)$};
   \draw[->,black] (5) to  (3);
\draw[->,black] (2) to  (4);
 \draw[->,black] (1) to  (3);
 \draw[->,black] (5) to  (1);
  \draw[->,black] (5b) to  (1);
   \draw[->,black] (5c) to  (1);
 \draw[->,black] (2) to  (3);
 \draw[->,black] (5) to  (3);
  \draw[->,black] (5b) to  (3);
   \draw[->,black] (5c) to  (3);
 \draw[->,black] (1) to  (4);
 \draw[->,black] (5b) to  (4);
  \draw[->,black] (5c) to  (4);
 \draw[->,black] (6) to  (3);
 
  \draw[->,black] (1) to  (2);
 
\end{tikzpicture} }
\vspace{5.2cm}
   
  \end{minipage}
\caption{\scriptsize{(a) The data generating mechanism in simulations. Panels (b) to (g) show the associations remaining after using the weights of the corresponding estimators (and boxes are used to represent the variables upon which we condition in each mean outcome model or, in the case of the observation indicator $dN(t)$, that we implicitly condition upon by using only observed data): (b) $\hat{\psi}_{DW1}$; (c) $\hat{\psi}_{DW2}$; (d) $\hat{\psi}_{DW3}$; (e) $\hat{\psi}_{DW4}$; (f) $\hat{\psi}_{IPT}$; and (g) $\hat{\psi}_{OLS}$. A dashed line represents a relationship that is possibly remaining due to a misspecified observation model (either an observation model lacking predictors, or for which some parameters are estimated with bias because of other dependent predictors missing in the model).} }

\end{figure}
 
Supplementary Figure 1 presents in (a) the data generating mechanism in simulation studies at time $t$. Note, the individual index is removed for ease of notation and interactions are not depicted in any diagrams in Supplementary Figure 1. Panels (b) to (g) show the associations remaining after using the weights of the corresponding estimators: (b) $\hat{\psi}_{DW1}$ (all models correctly specified); (c) $\hat{\psi}_{DW2}$ (partially misspecified observation model  w.r.t. $K_2$ and $K_3$, and misspecified outcome model w.r.t. $K_2$); (d) $\hat{\psi}_{DW3}$ (partially misspecified observation model w.r.t. $K_2$ and $K_3$, and misspecified treatment model w.r.t. $K_1$ and $K_3$); (e) $\hat{\psi}_{DW4}$ (misspecified observation model w.r.t. $Z(t)$ and $K_3$); (f) $\hat{\psi}_{IPT}$ (no adjustment for the observation model); (g) $\hat{\psi}_{OLS}$ (no adjustment for the observation model and no adjustment for the treatment model via an IPT weight). A box represents conditioning on the corresponding variable in the mean outcome model for all variables except the observation indicator $dN(t)$, which is implicitly conditioned upon by virtue of estimation relying only on observed data. A dashed line represents a relationship that is possibly remaining due to a misspecified model.
For figures (a), (e), (f) and (g) we find a path (an association) remaining that goes from $A(t)$ to $dN(t)$ to $Z(t)$ to $Y(t)$ that is not due to the causal effect of $A(t)$. The observation model adjusting only for $A(t)$ and $K_2$ is misspecified w r.t. to $Z(t)$ and $K_3$, but, as discussed in the main manuscript, it is also possibly misspecified with respect to $A(t)$ since that variable is associated with $Z(t)$. The coefficient for $A(t)$ in the observation model may, therefore, be biased in the subadjusted model containing only $A(t)$ and $K_2$.
\newpage

\noindent  \textbf{Supplementary Material B}\vspace{0.5cm}\\
\noindent \textbf{Simulation study results: error rate (i.e., empirical MSE) of the estimated optimal treatment decisions, absolute empirical bias of the blip values, absolute bias of each coefficient in the blip function, and average estimated value function, as obtained from the six alternative estimators} 
 
 \begin{table}[H]
\caption{Simulation study results ($M=1000$ simulations) for the comparison of error rate of the estimated optimal \textbf{treatment decision} obtained with six alternative models: DW1 the proposed doubly-weighted estimator which accounts for both processes correctly, DW2 for which the observation process was partially misspecified and the outcome model was misspecified, DW3 for which the treatment process was misspecified and the observation process was partially misspecified, DW4 for which the observation process was misspecified, OLS which does not adjust for confounding or observation process, and IPW which accounts only for confounding. Empirical MSEs are computed as the squared empirical bias of the estimated optimal treatment decision (based on the estimated blip function) plus its empirical variance.  The observation process varies but the confounding mechanism and the parameters of the true blip function remain the same in all 4 scenarios of varying $\bm{\gamma}$ below.}
\begin{center}
\begin{tabular}{ c c c  c  c c c c c }
 \hline
Sample&  $\bm{\gamma}^{\upsilon}$  &  No.~obs. times &  \multicolumn{6}{c}{Error rate}\\
size& parameters& mean (IQR)  &  $\hat{\bm{\psi}}_{DW1}$ &  $\hat{\bm{\psi}}_{DW2}$ & $\hat{\bm{\psi}}_{DW3}$ & $\hat{\bm{\psi}}_{DW4}$ & $\hat{\bm{\psi}}_{OLS}$ & $\hat{\bm{\psi}}_{IPT}$ \\
      \hline
     250 & 1 & 3 (1-3) & 0.02 & 0.01 & 0.01 & 0.04 & 0.03 & 0.04  \\ 
     
        ~ & 2 & 3 (2-5) & 0.05 & 0.06 & 0.05 & 0.16 & 0.15 & 0.16  \\  
        ~ & 3 & 6 (3-9) & 0.06 & 0.03 & 0.03 & 0.26 & 0.25 & 0.26  \\  
        ~ & 4 & 10 (8-12) & 0.01 & 0.01 & 0.00 & 0.01 & 0.00 & 0.01  \\ \hline
        500 & 1 & 3 (1-3) & 0.01 & 0.01 & 0.01 & 0.03 & 0.03 & 0.03  \\

        ~ & 2 & 3 (1-5) & 0.02 & 0.03 & 0.02 & 0.14 & 0.13 & 0.14  \\ 
        ~ & 3 & 6 (3-9) & 0.04 & 0.02 & 0.02 & 0.25 & 0.25 & 0.25  \\  
        ~ & 4 & 10 (8-12) & 0.00 & 0.00 & 0.00 & 0.00 & 0.00 & 0.00  \\  \hline
\end{tabular}
\end{center}
 \label{tab:char}
 
 \noindent \scriptsize{$\upsilon$.1. (-2, -0.3, 0.2, -1.2); 2. (0.3, -0.6, -0.4, -0.3); 3. (0.4, -0.8, 1, 0.6);  4.  (0, 0, 0, 0), i.e., uninformative observation.\\
Abbreviations: MSE, mean squared error; IQR, interquartile range.}

\end{table}
\newpage

\begin{table}[H]
\caption{Simulation study results ($M=1000$ simulations) for the comparison of \textbf{absolute bias} of the \textbf{blip values} obtained with six alternative models: DW1 the proposed doubly-weighted estimator which accounts for both processes correctly, DW2 for which the observation process was partially misspecified and the outcome model was misspecified, DW3 for which the treatment process was misspecified and the observation process was partially misspecified, DW4 for which the observation process was misspecified, OLS which does not adjust for confounding or observation process, and IPW which accounts only for confounding.  The observation process varies but the confounding mechanism and the parameters of the true blip function remain the same in all 4 scenarios of varying $\bm{\gamma}$ below.}
\begin{center}
\begin{tabular}{ c c c  c  c c c c c }
 \hline
Sample& $\bm{\gamma}^{\upsilon}$   &  Mean no.~obs. &  \multicolumn{6}{c}{Absolute bias}\\
size& parameters& times (IQR)  &  $\hat{\bm{\psi}}_{DW1}$ &  $\hat{\bm{\psi}}_{DW2}$ & $\hat{\bm{\psi}}_{DW3}$ & $\hat{\bm{\psi}}_{DW4}$ & $\hat{\bm{\psi}}_{OLS}$ & $\hat{\bm{\psi}}_{IPT}$ \\
      \hline
             250 & 1 & 3 (1-3) & 0.58 & 0.50 & 0.42 & 0.76 & 0.74 & 0.76  \\  
               ~ & 2& 3 (1-5) & 1.00 & 1.07 & 0.97 & 1.57 & 1.53 & 1.57  \\  
        ~ & 3 & 6 (3-9)  &1.14 &0.87 &0.83&2.04 &2.06&2.06   \\  
        ~ & 4 & 10 (8-12) & 0.24 & 0.24 & 0.21 & 0.24 & 0.19 & 0.24  \\ \hline
        500 & 1 & 3 (1-3) & 0.44 & 0.37 & 0.30 & 0.70 & 0.73 & 0.69  \\  
               ~ & 2 & 3 (1-5) & 0.73 & 0.80 & 0.70 & 1.54 & 1.53 & 1.54  \\ 
          ~ & 3 & 6 (3-9)  & 0.89&0.65& 0.63&2.04 &2.05 &2.05   \\  
        ~ & 4 & 10 (8-12) & 0.17 & 0.17 & 0.15 & 0.17 & 0.13 & 0.17  \\ \hline
        1000 & 1 & 3 (1-3) & 0.33 & 0.27 & 0.21 & 0.68 & 0.72 & 0.67  \\  
         
        ~ & 2 & 3 (1-5) & 0.55 & 0.60 & 0.53 & 1.50 & 1.49 & 1.50  \\  
         ~ & 3 & 6 (3-9)  & 0.69&0.48 &0.46&2.03 &2.04&2.04   \\  
        ~ & 4 & 10 (8-12) & 0.12 & 0.12 & 0.10 & 0.12 & 0.09 & 0.12  \\ \hline
        
        2500 & 1 & 3 (1-3) & 0.23 & 0.18 & 0.14 & 0.66 & 0.71 & 0.66  \\  
        
        ~ & 2 & 3 (2-5) & 0.36 & 0.40 & 0.35 & 1.51 & 1.51 & 1.50  \\  
           ~ & 3 & 6 (3-9)  & 0.52& 0.34 &0.32& 2.04 &2.05& 2.05  \\  
        ~ & 4 & 10 (8-12) & 0.08 & 0.08 & 0.07 & 0.08 & 0.06 & 0.08  \\ \hline
\end{tabular}
\end{center}
 \label{tab:char0bias}
 
 \noindent \scriptsize{$\upsilon$.1. (-2, -0.3, 0.2, -1.2);  2. (0.3, -0.6, -0.4, -0.3); 3. (0.4, -0.8, 1, 0.6); 4.  (0, 0, 0, 0), i.e., uninformative observation.\\
Abbreviations: MSE, mean squared error; IQR, interquartile range.}
\end{table}

\begin{longtable}{ c c c c c c c c }
 \caption{Simulation study results ($M=1000$ simulations) for the comparison of \textbf{absolute bias} of the \textbf{coefficients}  in the blip function obtained with six alternative models: DW1 the proposed doubly-weighted estimator which accounts for both processes correctly, DW2 for which the observation process was partially misspecified and the outcome model was misspecified, DW3 for which the treatment process was misspecified and the observation process was partially misspecified, DW4 for which the observation process was misspecified, OLS which does not adjust for confounding or observation process, and IPW which accounts only for confounding. The observation process varies but the confounding mechanism and the parameters of the true blip function remain the same in all 4 scenarios of varying $\bm{\gamma}$ below.}\\
  \hline 
 $\bm{\gamma}^{\upsilon}$ &  Estimator & \multicolumn{3}{c}{$n=250$} &  \multicolumn{3}{c}{$n=500$}\\
  parameters & for the ITR &  Intercept& $K_1$ & $Q$ & Intercept & $K_1$ & $Q$ \\ \hline
  1      & $\hat{\bm{\psi}}_{DW1}$  &0.09 &0.01 &0.03&0.09&0.01&0.02 \\
          & $\hat{\bm{\psi}}_{DW2}$  & 0.02& 0.01&0.04&0.05&0.00&0.02 \\
               & $\hat{\bm{\psi}}_{DW3}$  &0.02 &0.02 &0.02&0.02&0.00&0.01 \\
                    & $\hat{\bm{\psi}}_{DW4}$  & 0.69&0.00 &0.00&0.68&0.00&0.01 \\
                    & $\hat{\bm{\psi}}_{OLS}$  & 0.74 & 0.00 & 0.01 & 0.73 & 0.00 & 0.00 \\
     & $\hat{\bm{\psi}}_{IPT}$  &0.69 &0.00 &0.00&0.67&0.00&0.01 \\ \hline
  
                    2    & $\hat{\bm{\psi}}_{DW1}$  & 0.28&	0.01&	0.12&	0.13	&0.03	&0.09   \\
          & $\hat{\bm{\psi}}_{DW2}$  &0.37&	0.02	&0.12&	0.21&	0.05	&0.10\\
               & $\hat{\bm{\psi}}_{DW3}$  & 0.32&	0.02&	0.06&	0.14	&0.03&	0.07  \\
                    & $\hat{\bm{\psi}}_{DW4}$  & 1.51&	0.03&	0.01&	1.52	&0.01&	0.00  \\ 
                     & $\hat{\bm{\psi}}_{OLS}$  &1.50&	0.03	&0.01	&1.55	&0.01&	0.00   \\
     & $\hat{\bm{\psi}}_{IPT}$  &1.51	&0.03&	0.00	&1.52&	0.01&	0.01  \\\hline 
                     3     & $\hat{\bm{\psi}}_{DW1}$  & 0.50&	0.13&	0.13&	0.23	&0.05&	0.10  \\
          & $\hat{\bm{\psi}}_{DW2}$  & 0.22&	0.09&	0.09	&0.06&	0.01	&0.07  \\
               & $\hat{\bm{\psi}}_{DW3}$  &0.24	&0.09&	0.07&	0.05	&0.01&	0.07\\
                    & $\hat{\bm{\psi}}_{DW4}$  & 2.05&	0.00	&0.01	&2.03&	0.02&	0.00  \\ 
                    & $\hat{\bm{\psi}}_{OLS}$  &2.05&	0.01	&0.00&	2.03&	0.02	&0.00 \\
     & $\hat{\bm{\psi}}_{IPT}$  &2.06&	0.00	&0.01&	2.04	&0.01&	0.00 \\\hline
                   4      & $\hat{\bm{\psi}}_{DW1}$  & 0.00	&0.00&	0.00&	0.01	&0.01&	0.00 \\
          & $\hat{\bm{\psi}}_{DW2}$  & 0.00&	0.00	&0.00&0.01&	0.01	&0.00  \\
               & $\hat{\bm{\psi}}_{DW3}$  & 0.00&	0.01&	0.00&	0.01	&0.00	&0.00  \\
                    & $\hat{\bm{\psi}}_{DW4}$  &0.00&	0.00&	0.00&	0.01&	0.01&	0.00 \\
                    & $\hat{\bm{\psi}}_{OLS}$  &0.00&	0.01	&0.00&	0.01&	0.00	&0.00  \\
     & $\hat{\bm{\psi}}_{IPT}$  &0.00&	0.00&	0.00&	0.01	&0.01&	0.00  \\  \hline 
\end{longtable}
 \noindent \scriptsize{$\upsilon$.1. (-2, -0.3, 0.2, -1.2);  2. (0.3, -0.6, -0.4, -0.3);
  3. (0.4, -0.8, 1, 0.6);  4.  (0, 0, 0, 0), i.e., uninformative observation.}
  \normalsize

\begin{table}[H]
\caption{Simulation study results ($M=1000$ simulations, $n=25,000$) of the average estimated value function using the true data generating mechanism for all other variables than the treatment, and a treatment either based on the true data generating mechanism or on six alternative optimal treatment decisions: DW1 the proposed doubly-weighted estimator which accounts for both processes correctly, DW2 for which the observation process was partially misspecified and the outcome model was misspecified, DW3 for which the treatment process was misspecified and the observation process was partially misspecified, DW4 for which the observation process was misspecified, OLS which does not adjust for confounding or observation process, and IPW which accounts only for confounding. The observation process varies but the confounding mechanism and the parameters of the true blip function remain the same in all 4 scenarios of varying $\bm{\gamma}$ below. }
\begin{center}
\begin{tabular}{ c c c  ccccc}
 \hline 
      &   \multicolumn{7}{c}{Average estimated value function in the large dataset}        \\
 $\bm{\gamma}^{\upsilon}$        &  Actual treatment  & $\hat{\bm{\psi}}^{\dagger}_{DW1}$   & $\hat{\bm{\psi}}^{\dagger}_{DW2}$   & $\hat{\bm{\psi}}^{\dagger}_{DW3}$   & $\hat{\bm{\psi}}^{\dagger}_{DW4}$     &   $\hat{\bm{\psi}}^{\dagger}_{OLS}$    &$\hat{\bm{\psi}}^{\dagger}_{IPT}$     \\
      \hline
     1   &   -1.05  &0.54&0.55&0.55&0.52 & 0.53&0.52 \\
                                         2&   -2.86   &-0.31&-0.32&-0.30&-0.49&-0.45&-0.50  \\
                                       3 &    -3.82  &-1.34&-1.29&-1.29&-1.63&-1.61&-1.62  \\
                                            4  &    -0.86  &1.10&1.10&1.10&1.10 &1.10&1.10 \\ \hline
\end{tabular}
\end{center}
 \label{tab:char2}
\noindent \scriptsize{$\upsilon$. 1. (-2, -0.3, 0.2, -1.2); 2. (0.3, -0.6, -0.4, -0.3); 3. (0.4, -0.8, 1, 0.6);  4.  (0, 0, 0, 0), i.e., uninformative observation.\\
$\dagger$. Under the optimal treatment (as per the corresponding estimated optimal ITR).}\\
\end{table}

\normalsize

 \newpage
 
\noindent  \textbf{Supplementary Material C}\vspace{0.5cm}\\
\noindent \textbf{Flow chart in the application to the CPRD, United Kingdom, 1998-2017 } 
 
 \begin{figure}[H]
 \begin{center}
 \includegraphics[width=14cm]{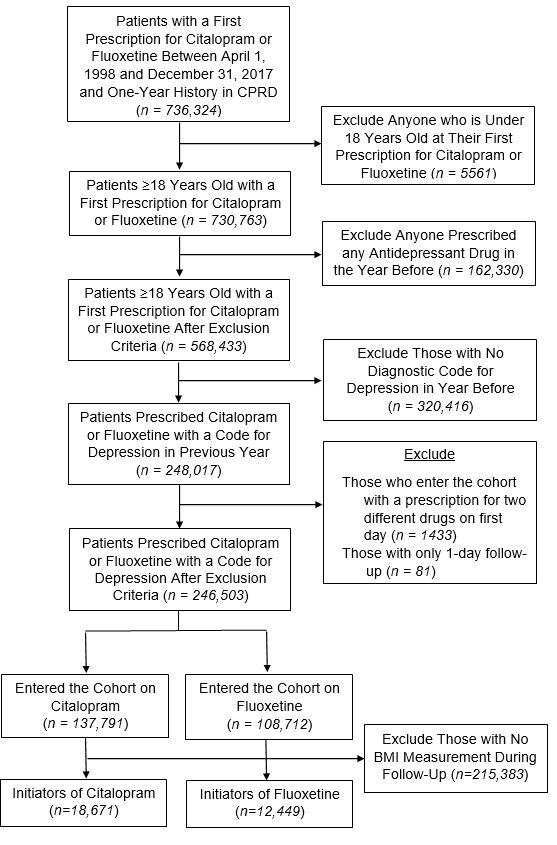}
 \end{center}
 \end{figure}

 \newpage
 
\noindent  \textbf{Supplementary Material D}\vspace{0.5cm}\\
  \noindent  \textbf{Baseline characteristics of the study cohort, observation rate ratios for the outcome, and estimated individualized treatment rules in the application to CPRD}\vspace{0.5cm}\\
 
 \begin{table}[H]
 \begin{center}
\caption{Baseline characteristics of the study cohort stratified by treatment at cohort entry ($n=31,120$), frequencies (\%, unless otherwise noted), CPRD, UK, 1998-2017 }

\begin{tabular}{lcc }
 \hline \hline
   &\multicolumn{2}{c}{Treatment} \\
Variable& Citalopram& Fluoxetine \\ 
 & (n=$18,671$)& (n=$12,449$)\\ \hline 
Age, mean (SD) &48.5 (18.1)&  45.1 (16.5) \\
 Male sex & 5965 (32) & 3609 (29)\\
 Index of Multiple Deprivation, mean (SD) & 3.0 (1.4)& 3.1 (1.4) \\
   Calendar year  & & \\
  \hspace{0.4cm} 1998-2005 & 3751 (20) & 4896 (39) \\
 \hspace{0.4cm} 2006-2011& 10,279 (55) & 5703 (46) \\
 \hspace{0.4cm} 2012-2017& 4641 (25)   & 1850 (15) \\
  Ever smoker & 11,586 (62) & 8017 (64) \\
Alcohol abuse &1478 (8)& 869 (7) \\
Psychiatric disease$^\dagger$ &521 (3)&321 (3) \\
Anxiety &5956 (32)&2987 (24) \\
Medication && \\
\hspace{0.4cm}  Antipsychotics  & 2836 (15)&1675 (13) \\
 \hspace{0.4cm}   Other psychotropic drugs$^\ddagger$  &4476 (24)& 2546 (20) \\
\hspace{0.4cm}    Lipid lowering drugs &3360 (18) & 1614 (13) \\
Number of psychiatric hospitalisations    & &\\
in previous 6 months, mean (SD)   & 0.04 (0.24)&  0.03 (0.34)\\
 \hline
\end{tabular}
 \label{tab:baseline}

 \scriptsize{ Abbreviations: CPRD, Clinical Practice Research Datalink; UK, United Kingdom; SD, standard deviation.\\
 $\dagger$. An indicator for a diagnosis of either autism spectrum disorder, obsessive compulsive disorder, bipolar disorder, or schizophrenia.\\
 $\ddagger$. Which include benzodiazepine drugs, anxiolytics, barbiturates and hypnotics.}
 \end{center}
 \end{table}

\begin{table}[H]
 \begin{center}
\caption{Estimated rate ratios (95\% bootstrap CIs) for the observation model, CPRD, UK, 1998-2017, n=$31,120$ individuals. }
\begin{tabular}{ l c}
 \hline  \hline
  & Rate ratio  \\ 
Variable& (Bootstrap 95\% CI) \\\hline
Antidepressant drug = citalopram  &  0.93  (0.91, 0.95)\\
Age& 1.00 (1.00, 1.00)\\
Male sex & 0.91   (0.89, 0.93) \\
Index of Multiple Deprivation & 1.03  (1.02, 1.03)\\
 Calendar year (Ref.= $<$2006) &\\
 \hspace{0.4cm} 2006-2011& 0.95 (0.92, 0.97) \\
 \hspace{0.4cm} 2012-2017&  0.91 (0.88, 0.93) \\
 Ever smoker & 1.69 (1.62, 1.70) \\      
Alcohol abuse &  1.01 (0.92, 1.07)\\
Psychiatric disease$^\dagger$ & 0.99  (0.86, 1.15)\\
Anxiety& 1.00  (0.97, 1.03) \\
Medication & \\ 
\hspace{0.4cm}Antipsychotics &  1.09 (1.02, 1.18)\\
\hspace{0.4cm}  Other psychotropic drugs$^\ddagger$  & 1.22 (1.17, 1.27)\\
\hspace{0.4cm} Lipid lowering drugs & 1.21 (1.17, 1.25)\\
Number of psychiatric hospitalisations in previous 6 months &  1.00 (0.95, 1.02)\\ 
\hline 
\end{tabular}
 \label{tabmonit1}
 
 \scriptsize{ Abbreviations: CI, confidence interval; CPRD, Clinical Practice Research Datalink; UK, United Kingdom; IMD, Index of Multiple Deprivation.\\
 $\dagger$. An indicator for a diagnosis of either autism spectrum disorder, obsessive compulsive disorder, bipolar disorder, or schizophrenia.\\
 $\ddagger$. Which include benzodiazepine drugs, anxiolytics, barbiturates and hypnotics.}
  \end{center}
 \end{table}

  \begin{landscape}
 
\begin{table}[H]
 \begin{center}
\caption{Coefficients of the blip function (95\% bootstrap CIs) for the optimal treatment rules as estimated by four alternative models: OLS which does not adjust for confounding or observation process, IPW which accounts only for confounding, IIV which accounts only for the observation process, and the proposed doubly-weighted estimator which accounts for both processes, CPRD, UK, 1998-2017, n=$31,120$ individuals. }
\begin{tabular}{ l c c c c } 
 \hline  \hline
Variable & $\hat{\bm{\psi}}_{OLS}$  & $\hat{\bm{\psi}}_{IPT}$  & $\hat{\bm{\psi}}_{IIV}$ & $\hat{\bm{\psi}}_{DW}$ \\ \hline
Intercept &-1.66 (-2.69, -0.46)&-1.38 (-2.62, -0.11)&-1.68 (-2.84, -0.58)& -1.45 (-2.66, -0.22)\\ 
Age &0.01 (-0.01, 0.03)&0.00 (-0.01, 0.03)&0.01 (-0.01, 0.03)& 0.00 (-0.02, 0.03)\\  
Male sex &-0.08 (-0.67, 0.55)&0.03 (-0.59, 0.62)&0.03 (-0.54, 0.65)& 0.16 (-0.48, 0.76)\\ 
IMD &0.14 (-0.09, 0.31)&0.14 (-0.13, 0.32)&0.12 (-0.10, 0.31)&0.13 (-0.13, 0.31)\\
Ever smoker &0.23 (-0.38, 0.66)&0.13 (-0.47, 0.65)&0.21 (-0.41, 0.66)& 0.08 (-0.50, 0.60)\\  
Alcohol abuse &1.03 (-0.11, 2.24)&0.68 (-0.48, 1.88)&0.78 (-0.26, 1.99)& 0.42 (-0.70, 1.60)\\
Psychiatric disease$^\dagger$ &0.44 (-1.83, 2.10)&1.02 (-1.23, 2.84)&0.58 (-1.73, 1.93)& 1.31 (-0.88, 3.05)\\  
Anxiety&0.29 (-0.07, 1.12)&0.31 (-0.02, 1.21)&0.32 (-0.05, 1.15)& 0.35 (0.00, 1.26)\\   
Medication & & & &\\
\hspace{0.2cm} Antipsychotics  &  -0.73 (-1.56, 0.17)&-0.82 (-1.75, 0.10)&-0.78 (-1.61, 0.10)&-0.91 (-1.91, 0.03)\\
 \hspace{0.2cm} Other psychotropic      &0.03 (-0.81, 0.66)&0.07 (-0.73, 0.64)&0.22 (-0.49, 0.86)& 0.30 (-0.47, 0.93)\\
 \hspace{0.4cm} drugs$^\ddagger$ &&&& \\
 \hspace{0.2cm} Lipid lowering drugs    & -0.16 (-0.73, 0.76)&0.04 (-0.70, 0.97)&-0.02 (-0.57, 0.95)&0.21 (-0.49, 1.23)\\
\hline
\end{tabular}
 \label{tab:appli}
 
 \scriptsize{ Abbreviations: CI, confidence interval; CPRD, Clinical Practice Research Datalink; UK, United Kingdom; IMD, Index of Multiple Deprivation.\\
 $\dagger$. An indicator for a diagnosis of either autism spectrum disorder, obsessive compulsive disorder, bipolar disorder, or schizophrenia.\\
 $\ddagger$. Which include benzodiazepine drugs, anxiolytics, barbiturates and hypnotics.}
   \end{center}
 \end{table}
 \end{landscape}
\newpage

\noindent  \textbf{Supplementary Material E}\vspace{0.5cm}\\

 \noindent \textbf{Blip function evaluated under different patient profiles of characteristics} \vspace{0.2cm}\\
 
 \begin{table}[H]
 \begin{center}  
\caption{Blip value under different patient profiles, CPRD, United Kingdom, 1998-2017 }
 \scriptsize{\begin{tabular}{c c c c c c c c c c }
 \hline \hline 
 Male &   & Ever   &Alcohol  &Psychiatric  & &Antipsy.&Psychotro.&Lipid& Value \\ 
 sex &  IMD &   smoker &  abuse&  diagnosis&Anxiety&drug&drug&lowering& blip \\
(yes) & (1 to 5) & (yes)  &(yes)&(yes)&(yes)&(yes)&(yes)&drug (yes)& function\\ \hline
0&1&0&0&0&0&1&0&0&-2.23 \\
1&1&0&0&0&0&1&0&0&-2.07 \\
0&3&0&0&0&0&1&1&1&-1.46 \\
0&1&0&0&0&0&0&0&0&-1.32 \\
1&3&0&0&0&0&1&1&1&-1.30 \\
1&1&0&0&0&0&0&0&0&-1.16 \\
0&3&1&0&0&0&0&0&0&-0.98 \\
1&3&1&0&0&0&0&0&0&-0.82 \\
0&5&0&0&0&0&0&0&0&-0.80 \\
0&3&0&0&0&1&0&0&0&-0.71 \\
1&5&0&0&0&0&0&0&0&-0.64 \\
0&3&1&1&0&0&0&0&0&-0.56 \\
1&3&0&0&0&1&0&0&0&-0.55 \\
1&3&1&1&0&0&0&0&0&-0.40 \\
0&5&1&1&0&0&0&0&0&-0.30 \\
1&5&1&1&0&0&0&0&0&-0.14 \\
0&3&0&1&1&0&0&0&0&0.67 \\
0&3&1&1&1&0&0&0&0&0.75 \\
1&3&0&1&1&0&0&0&0&0.83 \\
0&3&1&0&1&1&0&0&1&0.89 \\
1&3&1&1&1&0&0&0&0&0.91 \\
0&3&0&1&1&1&0&0&0&1.02 \\
0&5&1&1&1&0&0&0&0&1.01 \\
1&3&1&0&1&1&0&0&1&1.05 \\
1&5&1&1&1&0&0&0&0&1.17 \\
1&3&0&1&1&1&0&0&0&1.18 \\
1&5&1&1&1&1&0&1&1&2.03 \\
 \hline  
\end{tabular} }
 \label{tab:bb}

  \end{center}
 \end{table}
 
 \normalsize
Supplementary Table \ref{tab:bb} shows several profiles of individuals and the corresponding estimates of the blip function found using $\hat\psi_{DW}$ from our proposed approach. The sign of the estimated blip function indicates which treatment is to be recommended. For instance, a female with an Index of Multiple Deprivation of 1 who never smoked, had no alcohol abuse, no diagnosis for psychiatric diseases, no anxiety diagnosis, and who used antipsychotic drugs but did not use other psychotropic drugs or lipid-lowering drugs obtains the lowest blip value of -2.23 and, therefore, her recommended treatment is fluoxetine. A male with an Index of Multiple Deprivation of 5 who is an ever smoker, who had alcohol abuse, received a diagnosis for psychiatric disease, received a diagnosis for anxiety, did not use antipsychotic drugs but used other psychotropic drugs and lipid-lowering drugs obtains a blip value of 2.03 and, therefore, his recommended treatment is citalopram.
\end{singlespace}
 
\end{document}